\documentclass[aps,prb,amsmath,amssymb,twocolumn,letterpaper,%
         showpacs,balancelastpage,raggedbottom,%
         citeautoscript,floatfix,showkeys,superscriptaddress]{revtex4-2}
\usepackage{hyperref}
\hypersetup{
    pdftitle={Tuning magnitude and direction of lattice thermal conductivity in transition metal dichalcogenide heterobilayers},    
    pdfauthor={Elliot Perviz},     
    pdfkeywords={van der waals} {lattice thermal conductivity} {ab initio} {transition metal dichalcogenides} {design rules},
    colorlinks = true,
    citecolor  = ForestGreen,
    linkcolor  = RoyalBlue,
    urlcolor   = RoyalBlue
}
\usepackage{graphicx}
\usepackage[dvipsnames]{xcolor}
\usepackage{siunitx}
\usepackage{caption}
\usepackage{bm}
\usepackage{subcaption}
\usepackage{verbatim}

\usepackage{soul}

\begin{document}
\title{Tuning magnitude and direction of lattice thermal conductivity in transition metal dichalcogenide heterobilayers}

\author{Elliot Perviz}
\email{perviell@fel.cvut.cz}
\affiliation{Department of Control Engineering, Faculty of Electrical Engineering, Czech Technical University in Prague, Technicka 2, 16627, Prague 6, Czech Republic}

\author{Antonio Cammarata}
\email{cammaant@fel.cvut.cz}
\affiliation{Department of Control Engineering, Faculty of Electrical Engineering, Czech Technical University in Prague, Technicka 2, 16627, Prague 6, Czech Republic}

\date{\today}

\begin{abstract}
We investigate the nanoscale mechanisms determining lattice thermal conductivity (LTC) of pristine and W-doped MX$_2$–M$^\prime$X$^\prime_2$ transition metal dichalcogenide heterobilayers from first principles, using the exact solution of the linearised Boltzmann transport equation in both phonon and relaxon bases.
Pristine heterobilayers exhibit isotropic in-plane LTC with preserved ordering across temperature. 
Relaxon analysis identifies descriptors linking LTC to phonon properties such as the phonon group velocity and layer localisation.
While systems with lighter atoms generally favour higher LTC, a sufficiently large mass contrast is required to induce layer localisation of the transport-relevant vibrational modes.
Further, we show through the thermal viscosity that the relative distribution of vibrational states between metal/non-metal sublattices influences the balance between Normal and Umklapp scattering processes.
On the other hand, doped systems exhibit reduced and anisotropic in-plane LTC, retain a well-defined layer character, but are strongly affected by enhanced phonon-phonon scattering due to mass disorder.
Notably, we find that both configuration and temperature dictate the direction of maximum thermal transport, which opens the possibility to tune the direction of maximum (and minimum) conductivity via doping in novel 2D functional materials.
Thanks to its general formulation, the analysis protocol can be readily extended to other van der Waals heterostructures, and the descriptors may be implemented in high-throughput engines to identify promising layered materials with tailored thermal transport characteristics.
\end{abstract}

\maketitle

\section{Introduction}
\label{sec:intro}
Transition Metal Dichalcogenides (TMDs) are a versatile class of two-dimensional (2D) materials with tunable mechanical \cite{Bertolazzi2011, mos2-elasticity, tmd_mechprop} and electronic \cite{PhysRevLett.105.136805, Li2007} properties.
Advances in synthesis techniques \cite{Han2015} have enabled precise control over their structure, facilitating their integration into nanoscale devices spanning electronic \cite{Radisavljevic2011, Yoon2011}, optoelectronic \cite{Lopez-Sanchez2013, Frey2003}, electrochemical sensing \cite{mos2-glucose, mos2-NO-sensing}, energy storage \cite{B920277C, B820555H, Ratha2013}, and tribological \cite{ZHANG201967} applications.
In such applications, thermal transport plays a critical role.
Efficient heat dissipation is essential to mitigate the formation of localised hot spots, which can degrade device stability and performance in electronic and optoelectronic systems \cite{10.3389/fmats.2020.578791}.
Conversely, materials with intrinsically low thermal conductivity are desirable for thermoelectric applications, where suppressed heat transport enhances energy conversion efficiency \cite{https://doi.org/10.1002/adfm.201604134}.
In tribological contacts, frictional dissipation generates heat at sliding interfaces, and the ability of the material to conduct this away governs the interfacial temperature, which affects both wear and operational stability \cite{e12051021}.
These competing requirements make the control of thermal transport a central design parameter in TMD-based heterostructures.
A substantial body of work has characterised lattice thermal conductivity (LTC) in both monolayer and multilayer TMD systems.
First-principles Boltzmann transport equation (BTE) studies have shown that microscopic descriptors such as phonon group velocities, phonon band gaps, and the coupling between acoustic and low-frequency optical modes are key determinants of transport \cite{10.1063/1.4896685, PhysRevB.109.125422, ZHONG2020126676}.
In particular, large mass contrast within a single layer can induce phonon band gaps that suppress scattering channels and enhance phonon lifetimes \cite{10.1063/1.4896685}, and increasing thickness further modulates transport \cite{10.1063/1.4942827, PhysRevB.109.125422, SONG2018442}.
TMDs also exhibit a strong size effect due to reduced boundary scattering \cite{SONG2018442}, while isotope disorder can significantly affect the anharmonic interactions leading to modifications of the phonon linewidths and transport coefficients \cite{10.1063/1.4850995}.
In heterostructures, additional complexity arises from interlayer interactions.
Studies of TMD heterobilayers have demonstrated that the thermal conductivity of heterostructures often lies between those of the constituent monolayers, but can deviate substantially due to interlayer coupling, which introduces new scattering pathways involving low-frequency modes that enhance anharmonic phonon-phonon scattering \cite{C9CP01702J, C8RA10601K}.
Structural factors such as lattice mismatch and mass contrast between the layers influence the thermal conductivity in a non-trivial manner \cite{DEVRIES2023106447}, while symmetry breaking has been identified as a key mechanism for reducing thermal conductivity via increased scattering rates \cite{10.1063/5.0254641}.
Beyond composition and stacking, van der Waals heterostructures offer additional degrees of freedom for tuning thermal transport.
In particular, rotational disorder and twist angle engineering have emerged as powerful control parameters.
Random interlayer rotations can strongly suppress through-plane thermal conductivity and generate extreme thermal anisotropy \cite{Kim2021}, while periodic moiré structures enable systematic tuning of transport via modification of phonon dispersion and scattering phase space \cite{Eriksson2023, D2CP01304E}.
In TMD heterobilayers, phonon-phonon interactions can lead to significant mode coupling, such that thermal transport can no longer be interpreted in terms of independent phonon relaxations.
Even when anharmonic effects are fully included, the phonon basis does not diagonalise the scattering operator, and therefore does not provide a representation in terms of independent dissipative channels.
Instead, transport is more naturally described in terms of collective excitations of phonon populations, known as relaxons, defined as the eigenvectors of the scattering operator  \cite{PhysRevX.6.041013}.
This framework naturally captures the non-local and collective nature of heat transport arising from mode coupling.
To date, although numerous studies have systematically investigated the LTC of bilayer TMD heterostructures, a corresponding analysis within the relaxon framework remains unexplored.
This limits our ability to directly connect commonly used phonon-based descriptors to the underlying collective transport mechanisms.
In this work, we address this gap by providing a systematic relaxon-based analysis of thermal transport in TMD heterostructures.
We use quantitative descriptors such as the phonon group velocity and layer localisation of phonon modes to understand the microscopic origins of variation in the LTC, as well as probing probing the transport regime via the cophonicity \cite{cophonicity}.
Additionally, we investigate the LTC for systems beyond the ``pristine'' heterobilayer configuration (i.e. where each layer is a unique TMD) by considering W-doped Mo$_{1-x}$W$_{x}$S$_2$ for a selection of configurations with $x \in [0.0,1.0]$.
We demonstrate a pronounced reduction in thermal conductivity with doping, which is partially explained by the aforementioned descriptors.
Notably, we show that doping at different concentrations and distributions provides some level of control over the directional anisotropy of in-plane thermal conductivity in a temperature-dependent manner, providing a potential route for engineering thermal transport in low-dimensional functional devices.
%

\section{Computational details}
\label{sec:methods}
\begin{figure*}[t]
\centering
\includegraphics[width=\textwidth]{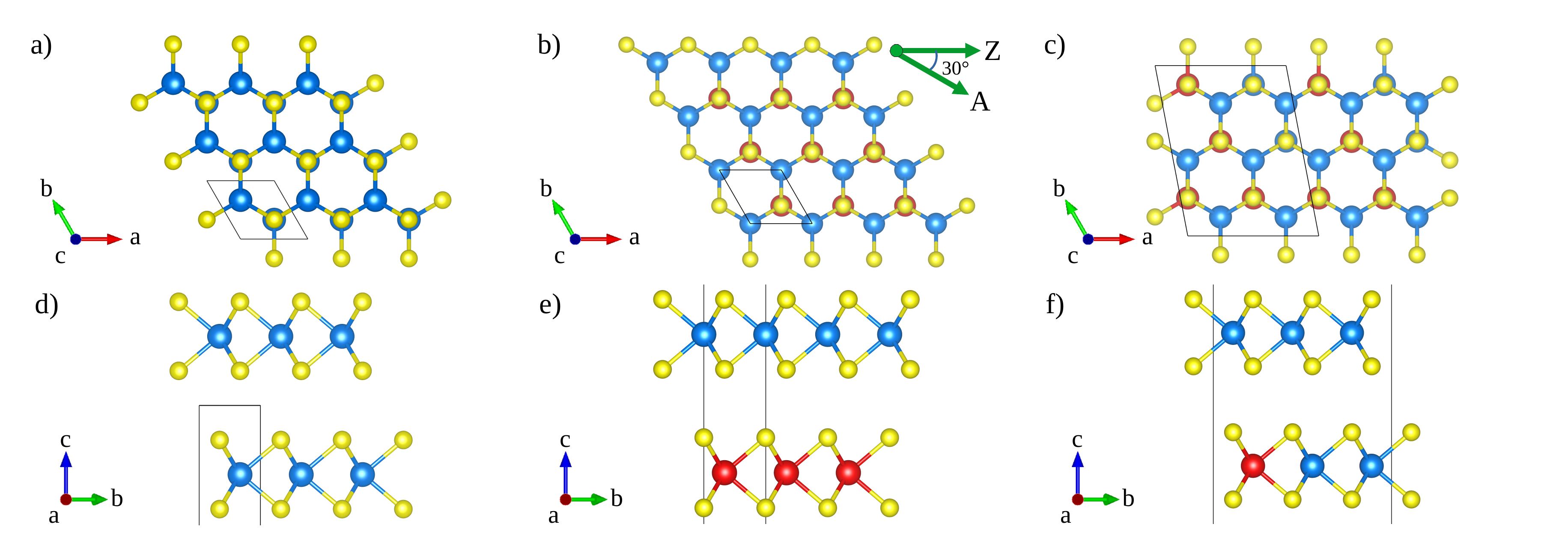}
\caption{Top-down views of a) homobilayer MoS$_2$, b) heterobilayer MoS$_2$-WS$_2$, and c) W-doped heterobilayer MoS$_2$-MoWS$_2$ (33\% W); corresponding side views are shown in d)-f), respectively.
Mo, S, and W atoms are depicted in blue, yellow, and red. 
Black lines in each plot draw the boundaries of the associated primitive cell.
Green arrows in b) indicate the $30^\circ$ sector between one pair of zigzag (Z) and armchair (A) crystallographic axes, where Z is fixed parallel to the $\bm{a}$ lattice vector and subsequently the Cartesian $x$ direction.}
\label{fig:struct}
\end{figure*}
Transition Metal Dichalcogenides (TMDs) are layered crystalline materials, built from repeating sequences of topologically flat planes bound together by weak van der Waals forces.
The general chemical formula of a single layer is MX$_2$, comprised of the transition metal M and two chalcogen atoms X.
In this study, we consider the 2H polymorph crystalline MX$_2$ compounds as reference structures \cite{Chhowalla2013} to build heterobilayers of the form MX$_2$-M$^\prime$X$_2^\prime$, where M,M$^\prime \in \{\text{Mo, W}\}$ and X,X$^\prime \in \{\text{S, Se, Te}\}$.
Excluding homobilayers and treating layer order as equivalent, this yields 15 unique systems with crystallographic space group P3$m$1 (no. 156).
In addition to this set of ``pristine'' heterobilayers, we consider W-doped Mo$_{1-x}$W$_{x}$S$_2$, for a selection of configurations with $x \in [0.0,1.0]$.
Starting from homobilayer MoS$_2$ ($x = 0.0$), W substitution is introduced within a single layer until we reach the MoS$_2$-WS$_2$ heterobilayer ($x = 0.5$); and subsequently within the second layer until we realise homobilayer WS$_2$ ($x = 1.0$).
We restrict our attention to doped heterostrucures which lie on the convex hull identified in our previous study \cite{PhysRevMaterials.8.106001}, thereby defining a physically relevant, non-random sequence of compositions.
For each system, the unit cell lattice parameters $\{a,b,c\}$ are defined as follows:
both MX$_2$ layers lie in the ($\bm{a},\bm{b}$) plane;
$a$ is aligned with the Cartesian $x$-direction;
$b$ lies in the ($\bm{x},\bm{y}$) plane at an angle of $120^{\circ}$ to $a$;
and $c$ is aligned with the Cartesian $z$-direction and fixed to $40$ \AA{}.
This choices provides a vacuum region perpendicular to the layer to suppress interactions between periodic images along this direction.
We calculate the optimised geometry of each system (reported in the Supporting Information section S1) using the Vienna Ab initio Simulation Package (\textsc{vasp}) \cite{VASP}, and subsequently use the interface between \textsc{vasp} and \textsc{phono3py} \cite{phono3py} to construct second (FC2) and third order (FC3) force constants by means of the finite displacement method, and finally calculate the lattice thermal conductivity (LTC) via solution of the linearised Boltzmann transport equation (LBTE) at varying temperature.
In all DFT calculations with \textsc{vasp} we use the Projector Augmented Wave approach to describe the atomic type \cite{PAW}, set exchange-correlation via the Perdew-Burke Ernzerhof Generalised Gradient Approximation \cite{pbe_gga}, and apply the DFT-D2 van der Waals correction in the Grimme formulation \cite{vdw-correction}.
We choose an energy cutoff of 500 eV for the plane-wave basis set while the density of the $k$-point mesh sampling is set according to the size of the system, at maximum $7 \times 7 \times 1$ for the primitive homobilayer primitive cells, and reduced accordingly depending on the size of the supercell.
The Self-Consistent Field procedure is converged up to a tolerance of $10^{-9}$ eV on the change in the total energy, and the maximum force component on any atom after geometry optimisation is $< 10^{-3}$ eV/\AA{}.
The above choices are made based on previous convergence studies in the literature on similar systems (see Refs. \citenum{antonio1, antonio2} and references therein).
Depending on the type of system, the specific finite displacement workflow used to construct the FC2 and FC3 tensors, and to solve the LBTE, differs.
For the pristine homobilayers and heterobilayers, the primitive cell contains 6 atoms, and we derive the required forces directly from ground state DFT calculations of finitely displaced ($3\times3\times1$) supercells.
We then calculate the LTC at varying temperature via exact solution of the LBTE \cite{PhysRevLett.110.265506} as implemented in \textsc{phono3py} on a ($19 \times 19 \times 1$) $q$-mesh, and perform single temperature analysis of the LTC at 300K by performing duplicate LBTE calculations in the relaxons framework \cite{PhysRevX.6.041013} using \textsc{phoebe} \cite{Cepellotti_2022}.
In \textsc{phono3py}, Brillouin zone integration is performed using the tetrahedron method, whereas \textsc{phoebe} employs a Gaussian smearing scheme, for which we find a converged broadening parameter of $0.0065$ eV to be sufficient for our systems.
On the other hand, the intermediate configurations of the Mo-W sulphide system have primitive cells containing 36 atoms.
Considering the fact that each intermediate configuration possesses P1 symmetry, the number of required finite displacements to build the third order force constants, which scales as $3N^2$, is 11,164 for the primitive cell and 186,624 for a ($2 \times 2 \times 1$) supercell.
Due to the computational expense of evaluating such a large number of DFT forces on the supercell, we derive force constants from the primitive cell due to its already large size (note: the doped primitive cells are originally made from $2 \times 2 \times 1$ supercells of the homobilayer reference).
Further, instead of directly calculating each of the finite displacements via DFT, we train a polynomial machine learning potential (MLP) \cite{pypolymlp1, pypolymlp2} for each intermediate configuration on a small subset of random finite displacements of the primitive cell, and thence utilise the MLP to evaluate the actual finite displacements required to construct the force constants systematically.
For each MLP, we checked the convergence of the total energy ($1 \times 10 ^{-6}$ eV), forces ($1 \times 10^{-3}$ eV/\AA{}), and lattice thermal conductivity ($0.5$ W/mK) derived from the exact solution of the LBTE from \textsc{phono3py} at varying temperature against the size of the subset of finite displacement DFT data.
A $19 \times 19 \times 1$ $q$-mesh is used consistently across all LBTE calculations, which is found to evaluate the LTC to sufficient convergence.
The LTC is obtained by solving the LBTE, which describes the response of the phonon population to a small temperature gradient.
The effect of phonon scattering enters through the scattering operator (collision matrix) $\bm{\Omega}$, which couples phonon modes across the Brillouin zone.
Within the single-mode relaxation time (SMRT) approximation, off-diagonal elements of $\bm{\Omega}$ are neglected.
Each component of the $3\times3$ LTC tensor, along directions $\{i,j\} \in \{x,y,z\}$, is then given by \cite{srivastava2022physics}
\begin{align}
 \kappa^{ij} &= \frac{1}{\mathcal{V}} \sum_\lambda \kappa_\lambda^{ij} = \frac{1}{\mathcal{V}} \sum_\lambda C_\lambda v_\lambda^i \Lambda_\lambda^{j}, \label{eq:rta_modal_decomposition} \\
 \Lambda_{\lambda}^{j} &= v_{\lambda}^j \tau_\lambda,
\end{align}
where $\mathcal{V}$ is the cell volume, $\lambda = (\bm{q}, s)$ labels the phonon mode at position $\bm{q}$ in reciprocal space and frequency branch $s$, $C_\lambda$ is the modal heat capacity, and $v_\lambda^i$ ($v_\lambda^j$) is the phonon group velocity component along the Cartesian direction $i$ ($j$).
In this approximation, phonons are treated as independent heat carriers, each characterised by a relaxation lifetime $\tau_\lambda$.
This neglect of mode coupling can lead to significant errors in systems where scattering is strongly collective \cite{PhysRevX.6.041013}.
The exact solution of the LBTE instead retains the full scattering operator and may be obtained either by directly solving the linear system in the phonon basis, as implemented in \textsc{phono3py} \cite{phono3py}, or by diagonalising the symmetrised scattering operator to work in a relaxon basis, as implemented in \textsc{phoebe} \cite{Cepellotti_2022}.
In the phonon basis beyond SMRT, transport is described in terms of coupled phonon modes, for which the quantities $\tau_\lambda$ no longer correspond to independent relaxation times due to mode coupling.
In contrast, diagonalisation of the scattering operator yields eigenvectors corresponding to collective excitations of phonon populations, known as \textit{relaxons}, with lifetimes given by the inverse eigenvalues.
In this basis, an arbitrary component of the $3 \times 3$ LTC tensor is expressed as \cite{PhysRevX.6.041013}
\begin{align}
  \kappa^{ij} &= C \sum_\alpha \kappa_\alpha^{ij} = C \sum_{\alpha} V^{i}_{\alpha} \Lambda^{j}_{\alpha}, \label{eq:relaxon_modal_decomposition} \\
  \Lambda_\alpha^{j} &= V_{\alpha}^{j} \tau_{\alpha}, \label{eq:relaxon_mfp}
\end{align}
where $\kappa^{ij}$ is the sum of relaxon mode contributions $\kappa_\alpha^{ij}$ indexed by $\alpha$.
Each contribution is constructed from the corresponding modal transport properties: the relaxon velocity $V_{\alpha}^{i}$, mean free path $\Lambda_\alpha^{i}$, and lifetime $\tau_\alpha$.
The total heat capacity is $C = (1/\mathcal{V}) \sum_\lambda C_\lambda$ with $C_\lambda$ the phonon modal heat capacities.
Consequently, thermal transport is described in terms of independent collective excitations with well-defined lifetimes, providing a natural decomposition into the underlying dissipative channels.
For TMD heterobilayers, the relaxon framework should be adapted to their quasi-two-dimensional nature.
In these systems, thermal transport is highly anisotropic, with cross-plane thermal conductivity strongly suppressed relative to the in-plane components, often by one to two orders of magnitude in layered materials \cite{Cai2023}.
We therefore restrict our analysis to the in-plane thermal transport, described by the $2 \times 2$ thermal conductivity tensor $\bm{\kappa}_{\text{2D}}$, with components $\kappa^{ij}$, $i,j \in \{x,y\}$.
Diagonalisation of $\bm{\kappa}_{\text{2D}}$ gives the principal conductivities $\{\kappa_{\text{max}}, \kappa_{\text{min}}\}$ and corresponding orthonormal eigenvectors $\{\bm{e}_{\text{max}}, \bm{e}_{\text{min}}\}$, which define the directions of maximal and minimal thermal transport.
We adopt $\kappa \equiv \kappa_{\text{max}}$ as a scalar measure of the maximum principal conductivity for comparison between systems.
Accordingly, modal transport quantities are projected onto $\bm{e}_{\text{max}}$,
\begin{equation}
 \label{eq:transport_property_projection}
 V_\alpha = \bm{V}_\alpha \cdot \bm{e}_{\text{max}}, \qquad \Lambda_\alpha = \bm{\Lambda}_\alpha \cdot \bm{e}_{\text{max}},
\end{equation}
where $\bm{V}_\alpha = (V_\alpha^x, V_\alpha^y)$ and $\bm{\Lambda}_\alpha = (\Lambda_\alpha^x, \Lambda_\alpha^y)$ denote the vector forms of the relaxon velocity and mean free path.
Projecting \autoref{eq:relaxon_modal_decomposition} then gives
\begin{equation}
 \label{eq:relaxon_mode_decomposition_proj}
 \kappa = C\sum_\alpha \kappa_\alpha = C \sum_\alpha V_\alpha \Lambda_\alpha.
\end{equation}
In isotropic systems, where $\kappa_{xx} = \kappa_{yy}$ and off-diagonal components vanish, the tensor is already diagonal so that $\kappa = \kappa_{\text{max}} = \kappa_{\text{min}}$, and $\kappa = \kappa^{xx} = \kappa^{yy}$.
Herein, when referencing the maximum/minimum principal conductivities they will simply be referred to as the maximum/minimum (in-plane) LTC ($\kappa_{\text{max}}$ / $\kappa_{\text{min}}$), and where transport is isotropic, simply as the (in-plane) LTC ($\kappa$).

In pristine heterobilayer TMD systems, the in-plane LTC tensor reflects the discrete rotational symmetry of the lattice, such that equivalent zigzag and armchair directions yield identical transport properties up to rotation.
By definition, the $a$ lattice parameter of the primitive hexagonal cell is aligned with one of the zigzag crystallographic axes, which we fix to be parallel with the Cartesian $x$ direction (see \autoref{fig:struct}), ensuring that comparisons between pristine systems are performed within a consistent crystallographic frame.
Doping breaks this rotational symmetry by introducing an inhomogeneous mass and force-constant distribution, and the resulting LTC tensor becomes dependent on the chosen coordinate frame.
To restore a consistent basis for comparison, we retain the same convention as for the pristine heterobilayers and align a zigzag direction with the $x$-axis, applying the corresponding rotation to the LTC tensor prior to analysis.
Diagonalisation of the rotated tensor yields $\{\kappa_{\text{max}}, \kappa_{\text{min}}\}$ and their associated eigenvectors $\{\bm{e}_{\text{max}}, \bm{e}_{\text{min}}\}$.
We then define an angular measure $\Delta \theta \in [0,30^{\circ}]$ as the smallest angle between $\bm{e}_{\text{max}}$ and the nearest zigzag direction.
This quantity is invariant under global rotations of the lattice and therefore provides a frame-independent measure of the anisotropy induced by doping.
Values of $\Delta \theta \approx 0^{\circ}$ and $\Delta \theta \approx 30^{\circ}$ correspond to transport aligned with zigzag and armchair directions, respectively, while intermediate values indicate a rotation of the preferred heat-flow direction away from the crystallographic axes.
The orthogonal eigenvector $\bm{e}_{\text{min}}$ defines the direction of minimal conductivity, with associated scalar $\kappa_{\text{min}}$, and is rotated by $90^{\circ}$ relative to $\bm{e}_{\text{max}}$.
In the pristine limit, the in-plane LTC tensor is isotropic, the eigenvalues are degenerate, and $\Delta \theta=0$, reflecting the absence of a preferred in-pane transport direction.
Under our convention, the Cartesian axes coincide with zigzag and armchair crystallographic axes, which are therefore equivalent.
Now, in the analysis that follows, we will regularly employ transport-weighted modal averages of relaxon/phonon modal quantities (e.g. $V_\alpha$, $\tau_\alpha$, $\Lambda_\alpha$; $v_\lambda$, $\tau_\lambda$, $\Lambda_\lambda$).
For a general transport property $A$, we define
\begin{equation}
\label{eq:transport_weighting}
 \left< A \right>_\kappa = \frac{\sum_{m} \kappa_{m} A_m}{\sum_m \kappa_{m}}, \quad m \in \{\alpha,\lambda\},
\end{equation}
where the summation is performed over either relaxon modes ($m=\alpha$) or phonon modes ($m=\lambda$), depending on the representation.
This weighting assigns greater importance to modes that dominate the thermal conductivity.
%

\section{Results}
\label{sec:results}

\subsection{Pristine heterobilayers}
\label{sec:prist_hetbi}

\begin{figure}[t]
\centering
\includegraphics[width=0.48\textwidth]{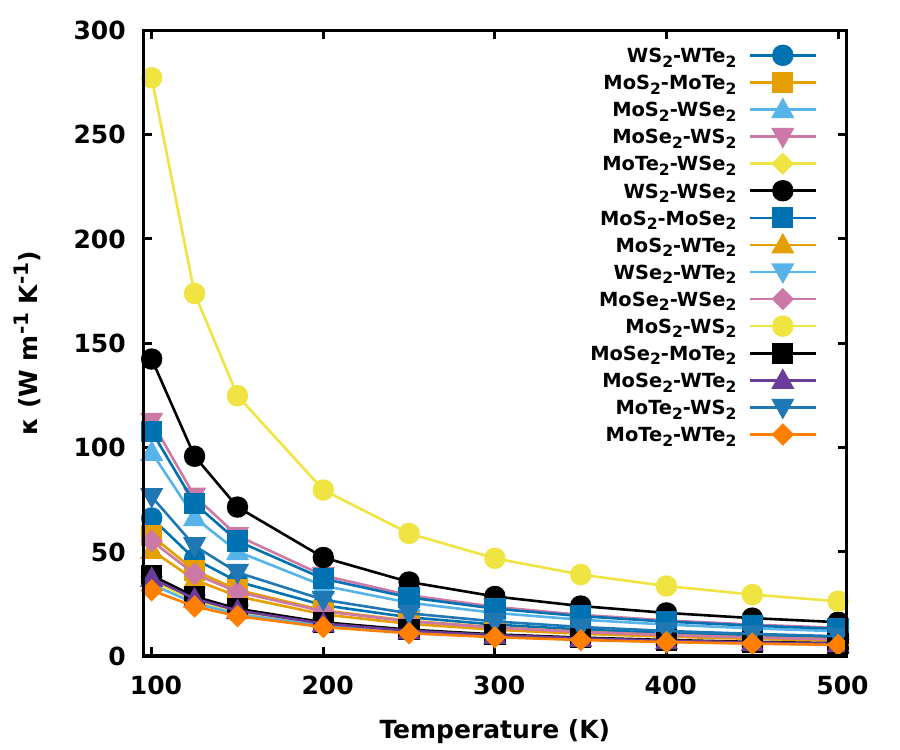}
\caption{Maximum principal conductivity derived from the exact solution of the LBTE via \cite{phono3py} as a function of temperature for the set of unique pristine MX$_2$ heterobilayers.}
\label{fig:ltc_vs_temp}
\end{figure}
\begin{figure*}[t]
\centering
\begin{minipage}{0.49\textwidth}
    \centering
    \includegraphics[width=\linewidth]{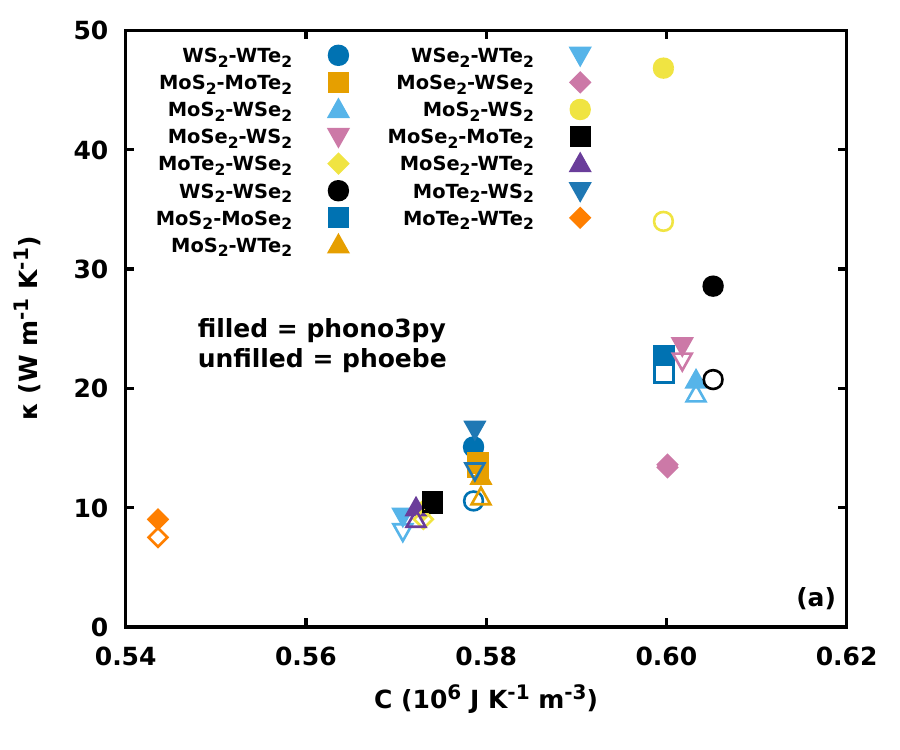}
\end{minipage}
\hfill
\begin{minipage}{0.49\textwidth}
    \centering
    \includegraphics[width=\linewidth]{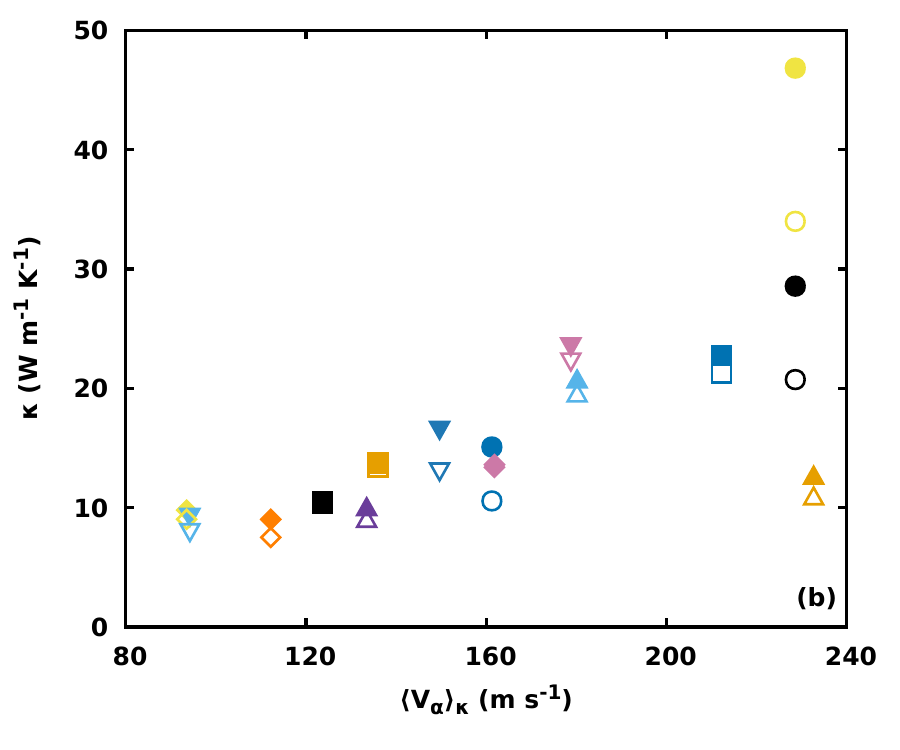}
\end{minipage}

\vspace{0.5em}

\begin{minipage}{0.49\textwidth}
    \centering
    \includegraphics[width=\linewidth]{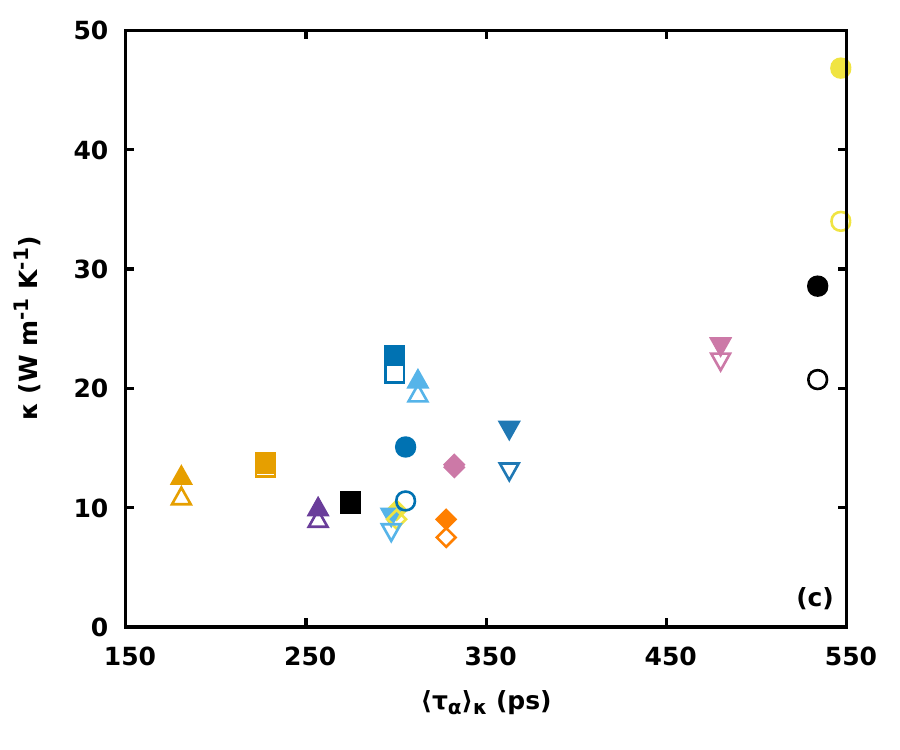}
\end{minipage}
\hfill
\begin{minipage}{0.49\textwidth}
    \centering
    \includegraphics[width=\linewidth]{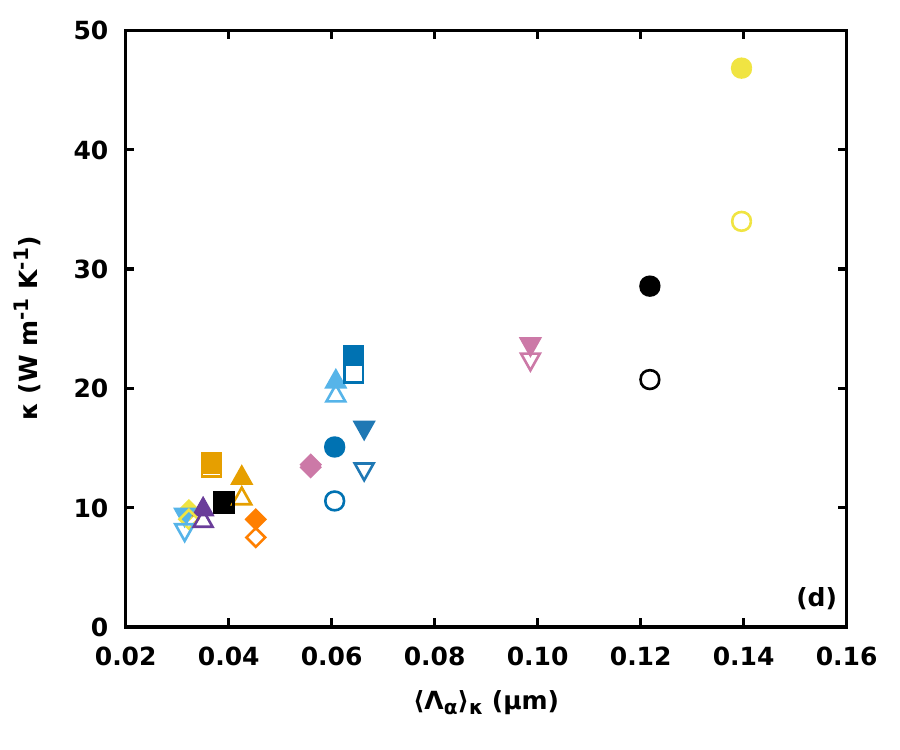}
\end{minipage}
\caption{
Maximum (in=plane) conductivities ($\kappa = \kappa_{\text{max}} = \kappa_{\text{min}}$) for the specified set of TMD heterobilayers, as a function of:
(a) heat capacity $C$;
(b) transport-weighted modal average relaxon velocities $\left<V_\alpha\right>_\kappa$;
(c) transport-weighted modal average relaxon lifetimes $\left<\tau_\alpha\right>_\kappa$;
(d) transport-weighted modal average relaxon mean free paths $\left<\Lambda_\alpha\right>_\kappa$.
Filled and unfilled markers correspond to results obtained from \textsc{phono3py} and \textsc{phoebe}, respectively.
}
\label{fig:ltc_vs_relaxon_properties}
\end{figure*}
\begin{figure*}[t]
\centering

\includegraphics[width=\linewidth]{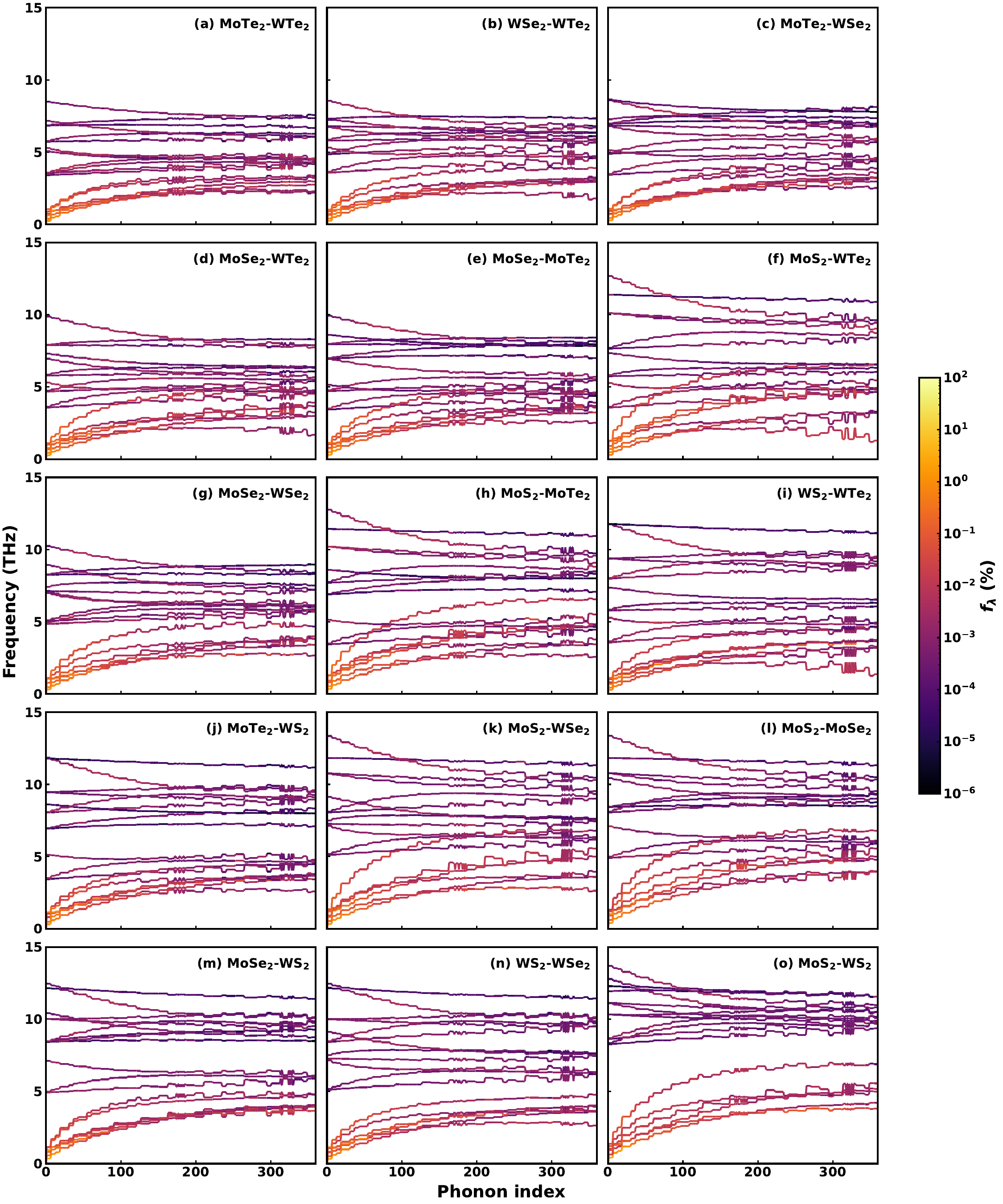}

\caption{$|\bm{q}|$-sorted phonon frequencies sampled on the transport mesh, with systems ordered from smallest to largest LTC in plots (a)-(o), determined via \textsc{phono3py} LBTE calculations.
The colour distinguishes the percentage contribution of each phonon mode (expressed in \%) to the the maximum (in-plane) LTC.}
\label{fig:relaxon_grid}
\end{figure*}
\begin{figure*}[t]
\centering
\begin{minipage}{0.49\textwidth}
    \centering
    \includegraphics[width=\linewidth]{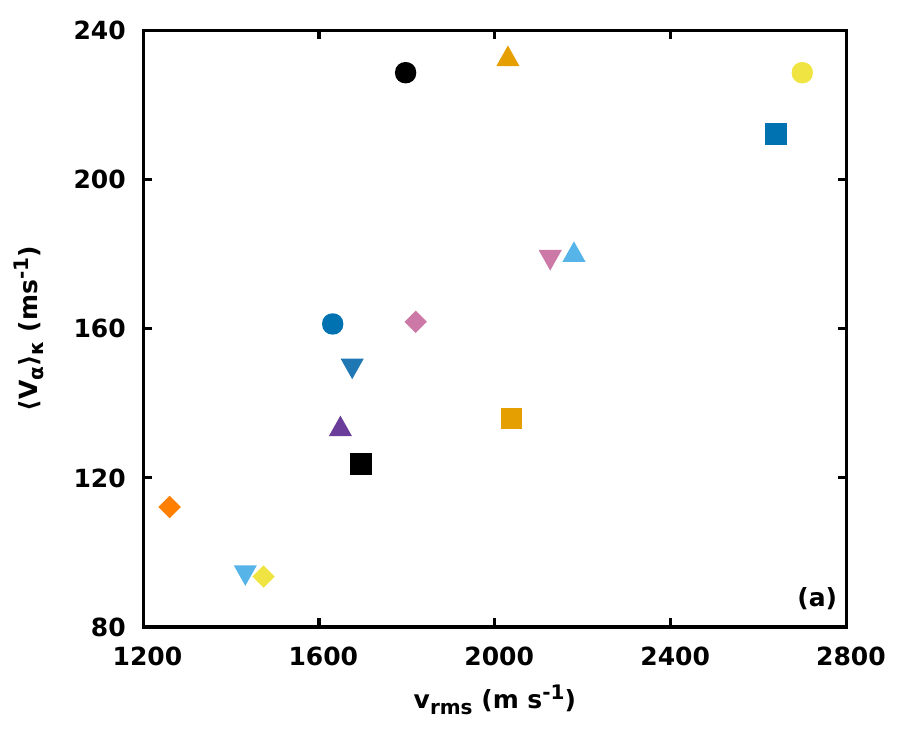}
\end{minipage}
\hfill
\begin{minipage}{0.49\textwidth}
    \centering
    \includegraphics[width=\linewidth]{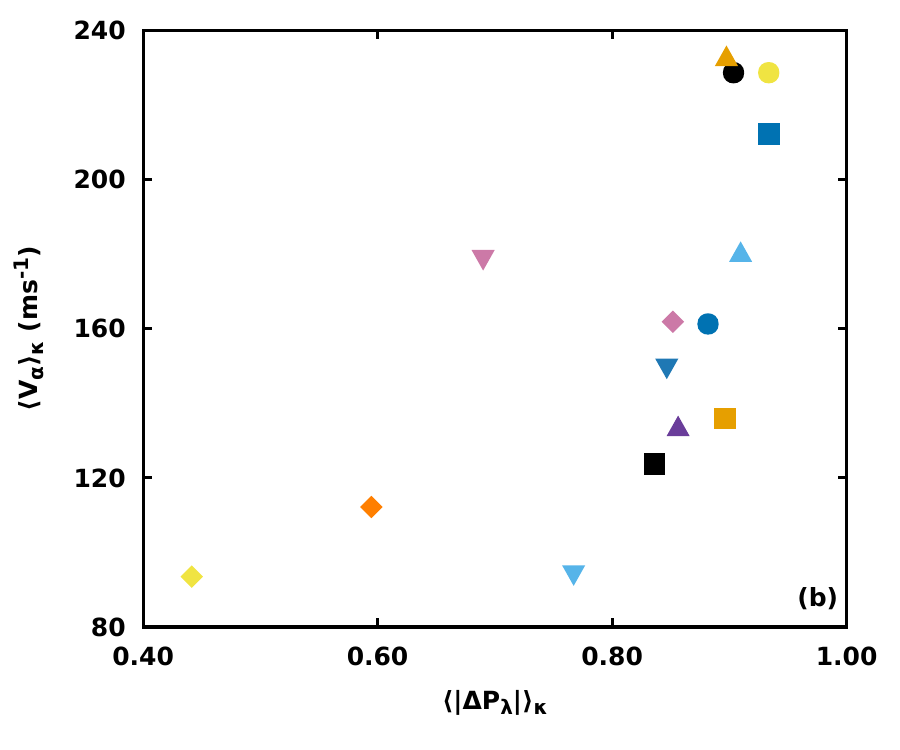}
\end{minipage}

\vspace{0.5em}

\begin{minipage}{0.49\textwidth}
    \centering
    \includegraphics[width=\linewidth]{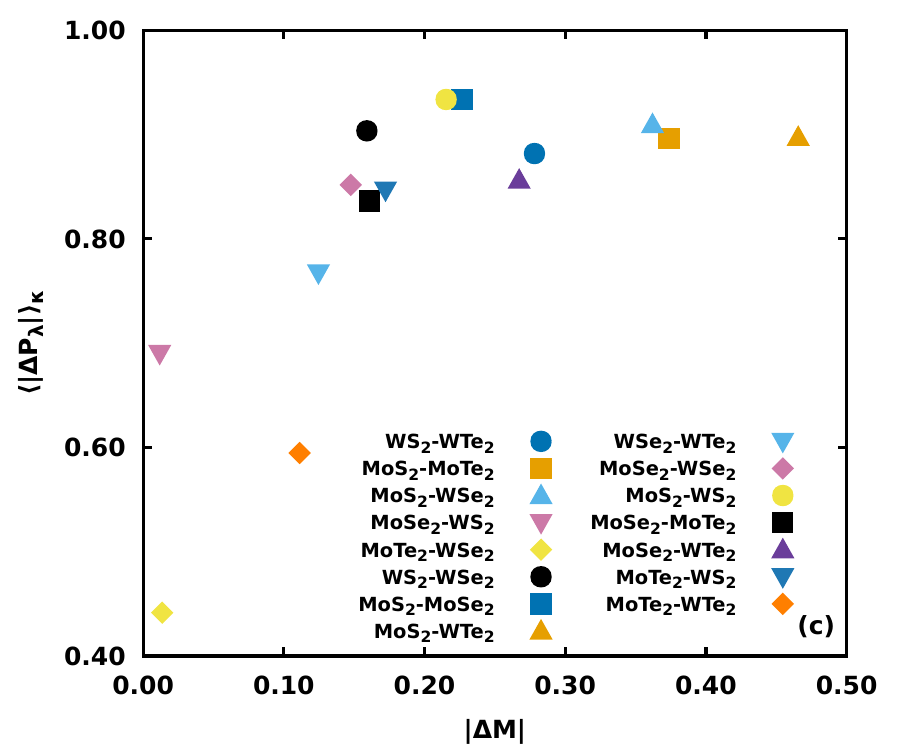}
\end{minipage}
\hfill
\begin{minipage}{0.49\textwidth}
    \centering
    \includegraphics[width=\linewidth]{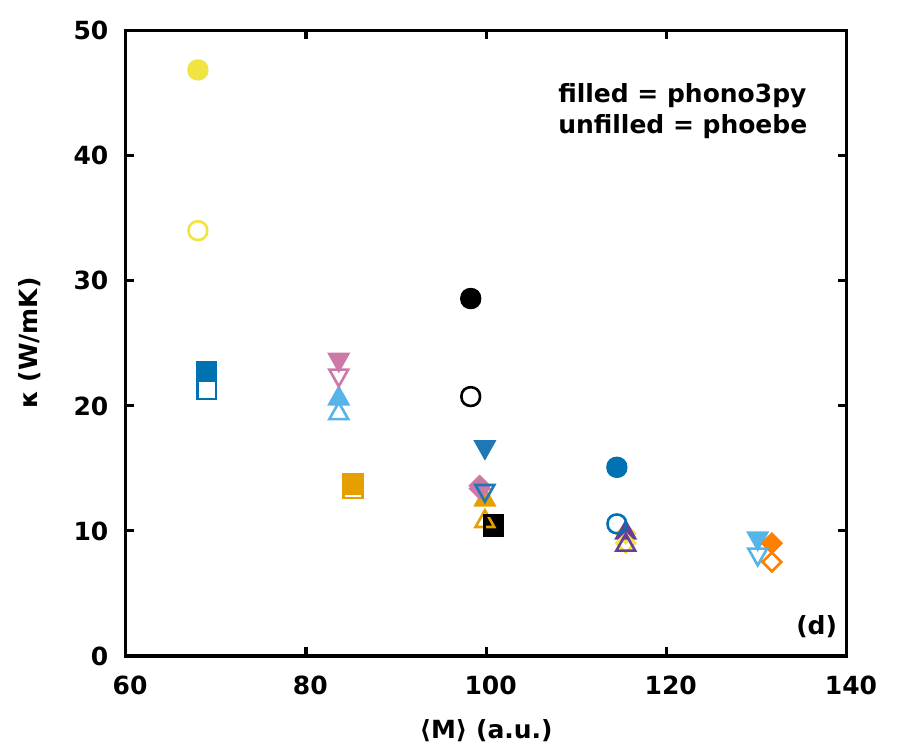}
\end{minipage}
\caption{
(a) Transport weighted modal average relaxon velocity $\left<V_\alpha\right>_\kappa$ plotted against the root mean square average phonon group velocities $v_{\text{rms}}$ in the 0-4 THz frequency range.
(b) $\left<V_\alpha\right>_\kappa$ as a function of the transport-weighted modal average phonon layer participation $\left<\Delta P_\lambda\right>_\kappa$.
(c) $\left<\Delta P_\lambda\right>_\kappa$ as a function of the layer mass contrast $|\Delta M|$.
(d) Lattice thermal conductivity $\kappa$ as a function of the average mass $\left<M\right>$.
}
\label{fig:deltaP_combined}
\end{figure*}

We calculated the LTC for the set of pristine TMD heterobilayers in the range $[100,500]$ K via exact solution of the LBTE in a phonon basis using \textsc{phono3py}.
Thermal transport is observed to be isotropic, and decreases as a function of temperature (\autoref{fig:ltc_vs_temp}), which reflects the well-known phenomenon of enhanced phonon-phonon scattering at elevated temperatures.
Further, since none of the curves intersect, we assess that there is a temperature independent ordering of the heterobilayers.
While these trends characterise the overall magnitude of thermal transport, they do not by themselves provide insight into the microscopic mechanisms governing the differences between systems.
In this context, previous work by Farris et al. \cite{PhysRevB.109.125422} demonstrated that key features of the phonon band structure --- such as the magnitude of group velocities ($|\bm{v}_g|$) near the $\Gamma$-point, the height of the frequency plateau in the $2$-$4$ THz frequency range, and the presence of a frequency gap --- can be used to rationalise relative trends in LTC across TMD homobilayers.
Given the structural similarities between those systems and the heterobilayers considered here, similar correlations are expected to emerge.
To establish a quantitative connection between these band-structure descriptors and thermal transport, it is necessary to resolve the LTC into its underlying modal contributions.
As discussed in \autoref{sec:methods}, this can be achieved via the relaxon representation \cite{PhysRevX.6.041013}, which provides a decomposition of the conductivity into independent dissipative channels.
Given that the ordering of systems by LTC is preserved across temperature, we perform relaxon BTE calculations at a representative temperature of 300 K.
We then examine how the LTC correlates with transport-weighted averages of relaxon mode properties, namely velocities $\left<V_\alpha\right>_\kappa$, lifetimes $\left<\tau_\alpha\right>_\kappa$ and mean free paths $\left<\Lambda_\alpha\right>_\kappa$, by comparing results from \textsc{phono3py} and \textsc{phoebe} (\autoref{fig:ltc_vs_relaxon_properties}).
It is evident that, on a case by case basis, there are non-negligible quantitative differences in the LTC between the two codes.
The largest differences are seen for the two systems with the largest LTC: MoS$-2$-WS$_2$ and WS$_2$-WSe$_2$.
Previous benchmarking studies \cite{phonon_olympics} have demonstrated that variations in computed LTC can arise from both methodological choices (e.g. Brillouin zone sampling, supercell size) and intrinsic differences in numerical implementation.
In the present case, where identical second- and third-order force constants are employed, the differences between \textsc{phono3py} and \textsc{phoebe} can be attributed primarily to two factors.
First, the treatment of the Brillouin zone integration differs: \textsc{phono3py} employs the tetrahedron method, whereas \textsc{Phoebe} uses Gaussian smearing.
It is well established that Gaussian smearing can broaden energy-conserving processes and may fail to accurately resolve sharp spectral features unless very dense $q$-point meshes are used  \cite{TORIYAMA2022100002}.
Second, the two approaches differ in how the scattering operator is represented and solved:
\textsc{phono3py} operates in a phonon basis, while \textsc{phoebe} uses relaxons.
Despite these differences, both approaches exhibit consistent correlations between LTC and relaxon transport properties as shown in \autoref{fig:ltc_vs_relaxon_properties}, which validates the use of \textsc{phono3py} for reference conductivity values and \textsc{phoebe} for analysing the microscopic transport behaviour in the relaxon basis.
The interpretation of \autoref{fig:ltc_vs_relaxon_properties} is then straightforward:
First, in \autoref{fig:ltc_vs_relaxon_properties}a, we plot the LTC against the volumetric heat capacity $C$.
A larger $C$ generally correlates with larger LTC, but there is substantial variability at fixed $C$, limiting its predictive utility.
Second, in \autoref{fig:ltc_vs_relaxon_properties}b we plot the LTC against the transport-weighted modal average relaxon velocity, $\left<V_\alpha\right>_\kappa$, which exhibits an approximately linear correlation.
MoS$_2$-WTe$_2$ is, however, a clear outlier, having high $\left<V_\alpha\right>_\kappa$ but low LTC.
In \autoref{fig:ltc_vs_relaxon_properties}c, we plot the LTC against the transport-weighted modal average relaxon lifetime, $\left<\tau_\alpha\right>_\kappa$.
The correlation is weaker, although a general trend remains whereby larger lifetimes correspond to higher LTC.
These results are expected, as the LTC emerges from the coupling of the relaxon mode properties above (recall \autoref{eq:relaxon_modal_decomposition}), such that descriptors probing only a single quantity cannot fully capture its variability, although they still produce rough trends.
By contrast, when the LTC is plotted against the transport-weighted modal average mean free path (MFP) (\autoref{fig:ltc_vs_relaxon_properties}d), a strong linear correlation is observed.
This reflects the fact that the MFP encodes the joint dependence on relaxon velocities and lifetimes, which sets the characteristic length scale of thermal transport.
Further, when expressed in terms of the MFP, the previously identified outlier is resolved; all systems follow a consistent trend with no signficant outliers.
Having investigated the underlying transport mechanisms, we now examine the relationship between the relaxons and the phonon basis in which they are defined.
Each relaxon eigenvector $\theta_\alpha$ spans the full phonon space indexed by $\lambda = (\bm{q}, s)$, such that $\theta_\alpha^{(\lambda)}$ quantifies the participation of phonon mode $\lambda$ in relaxon $\alpha$.
The contribution of a given phonon mode to the total conductivity can then be derived by projecting the relaxon modal decomposition onto the phonon basis:
\begin{equation}
 \label{eq:relaxon_bs_weight}
 \kappa_\lambda = \sum_\alpha \kappa_\alpha |\theta_\alpha^{(\lambda)}|^2,
\end{equation}
with the corresponding fractional contribution $f_\lambda = \kappa_\lambda / \kappa$
Since phonon frequencies associated at each $\bm{q}$-point in the $19 \times 19 \times 1$ transport mesh are obtained as a by-product of solving the LBTE, we can exploit \autoref{eq:relaxon_bs_weight} to construct the transport-weighted band structure for each TMD heterobilayer (\autoref{fig:relaxon_grid}).
We sort the $\bm{q}$-points according to their magnitude $|\bm{q}|$, where $\Gamma \equiv |\bm{q}| = 0$, to provide a consistent ordering across all systems; this means that consecutive points on the plot may correspond to physically separated regions of the of the Brillouin zone.
The band structures are ordered from smallest to largest LTC, following the temperature-independent ranking identified earlier (\autoref{fig:ltc_vs_temp}).
In all systems, the LTC is dominated by low-frequency phonon modes ($< 2$ THz) near $\Gamma$, with smaller contributions from modes further away; higher-frequency optical modes contribute negligibly.
The dominant contributions arise from the three acoustic branches and the three lowest frequency optical branches, which we collectively refer to as the ``low-frequency'' branches.
These branches originate near $\Gamma$ at $\approx 0$ THz and increase in frequency with $\bm{q}$ until reaching a plateau.
We may identify several qualitative correlations betwen the LTC and features of the band structure.
First, the LTC shows an increasing trend with the average frequency plateau of the low-frequency branches.
Second, the LTC correlates positively with the presence and magnitude of a frequency mode gap separating low-frequency branches from higher-frequency optical modes.
Third, the LTC generally increases with the overall phonon frequency range.
Due to the fact that the sequence of $|\bm{q}|$-points may be physically separated in reciprocal space, it is not possible to infer trends in phonon group velocities directly based on the curvature of the band structure reported in \autoref{fig:relaxon_grid}, as we might for a standard piece-wise linear path.
Instead, we perform a quantitative analysis using the phonon group velocities computed on the $19 \times 19 \times 1$ transport mesh.
Focusing on the modes which dominate the LTC, \autoref{fig:deltaP_combined}a shows $\left<V_\alpha\right>_\kappa$ plotted against the root mean square average phonon group velocity $v_{\text{rms}}$, projected along $\bm{e}_{\text{max}}$ in the $[0,4]$ THz frequency range, where we observe an approximate positive linear correlation.
Given the relaxon velocities display a near-linear scaling with the LTC (recall \autoref{fig:ltc_vs_relaxon_properties}a), this suggests that, at a coarse level, systems with larger $v_{\text{rms}}$ in the $[0,4]$ THz range tend to exhibit higher LTC, in agreement with previous observations \cite{PhysRevB.109.125422}.
$v_{\text{rms}}$ captures the typical magnitude of phonon group velocities in $[0,4]$ THz, however, it neglects the transport weighting encoded in the phonon or relaxon BTE.
As such, deviations from the observed trend are expected.
In particular, MoS$_2$-WTe$_2$, previously identified as an outlier in \autoref{fig:ltc_vs_relaxon_properties}, exhibits smaller $v_{\text{rms}}$ than would be suggested by its $\left<V_\alpha\right>_\kappa$.
This behaviour can be understood if we consider the transport-weighted band structure of \autoref{fig:ltc_vs_relaxon_properties}f, where significant contributions to the LTC arise from modes outside the $[0,4]$ THz.
These contributions are captured in $\langle V_\alpha \rangle_\kappa$ but are excluded from the definition of $v_{\text{rms}}$, leading to the apparent discrepancy.
A similar argument applies to WS$_2$-WSe$_2$, where $v_{\text{rms}}$ underestimates the effective transport velocity.
Inspection of \autoref{fig:ltc_vs_relaxon_properties}n shows that the transport weight is strongly concentrated in acoustic modes near $\Gamma$, at frequencies well below 4 THz.
In this case, $v_{\text{rms}}$ is diluted by modes which do not significantly participate in transport.
Overall, while  $v_{\text{rms}}$ provides a useful indicator of trends, it is a simplified description of the underlying transport physics, and thus should not be used alone for quantitative predictions of the LTC.
To further investigate the origins of the LTC trends, we now consider the layer character of the heat-carrying phonon modes.
In TMD heterobilayers, where the constituent layers exhibit distinct phonon spectra, the degree of interlayer coupling is expected to influence thermal transport.
We therefore define the modal layer participation factor as
\begin{equation}
 P_\lambda^{(\ell)} = \sum_{i \in \ell} \left| \mathbf{e}_{i\lambda} \right|^2, 
\end{equation}
i.e. the layer-resolved eigenvector norm, where $\mathbf{e}_{i\lambda}$ is the phonon eigenvector component associated with atom $i$ in mode $\lambda = (\mathbf{q}, s)$, and $\ell \in \{1,2\}$ labels the layer.
We then define the layer localisation of each mode as
\begin{equation}
 \label{eq:layer_localisation}
 |\Delta P_\lambda| = \left| P_\lambda^{(1)} - P_\lambda^{(2)} \right|
\end{equation}
where a value close to zero describes a phonon mode with roughly equal layer contributions, whereas a value close to 1 corresponds to a phonon mode almost completely localised on a single layer.
We evaluate the corresponding transport-weighted phonon mode average $\left< |\Delta P| \right>_\kappa$ according to \autoref{eq:transport_weighting}, which quantifies the extent to which heat-carrying phonon modes are preferentially localised on any one layer.
If we plot $\left<V_\alpha\right>_\kappa$ as a function of $\left<|\Delta P_\lambda|\right>_\kappa$ (\autoref{fig:deltaP_combined}b), we observe that the relaxon velocity of the modes contributing most strongly to transport increases approximately exponentially as the vibrational character becomes increasingly localised on a single layer.
The correlation between $\left<V_\alpha\right>_\kappa$ and $\kappa$ (\autoref{fig:ltc_vs_relaxon_properties}a) implies a similar relationship between  $\left<|\Delta P_\lambda|\right>$ and the LTC.
Importantly, $\left<|\Delta P_\lambda|\right>$ is itself correlated with the layer mass contrast,
\begin{equation}
 |\Delta M| = \left|\frac{\left<M^{(1)}\right> - \left<M^{2}\right>}{\left<M^{(1)}\right> + \left<M^{2}\right>}\right|
\end{equation}
where $\left<M^{(\ell)}\right>$ is the average mass of layer $\ell \in \{1,2\}$.
The quantity $|\Delta M|$ provides a normalised measure of the mass contrast between the two layers, with $\Delta M \rightarrow 0$ for identical layers and $|\Delta M| \rightarrow 1$ for strongly mismatched compositions.
As shown in \autoref{fig:deltaP_combined}c, $\left<|\Delta P_\lambda|\right>$ increases systematically with increasing $|\Delta M|$ before plateauing at $\left<|\Delta P_\lambda|\right> \approx 0.9$ beyond $|\Delta M| \approx 0.2$, at  which point further increasing $\Delta M$ does not strongly affect $\left<|\Delta P_\lambda|\right>$.
This suggests that to increase the LTC, a larger $|\Delta M|$ is a necessary initial condition (affecting the lower LTC range), but does not delineate the apparent exponential increase in $\left<V_\alpha\right>_\kappa$ as $\left<|\Delta P_\lambda|\right>$ approaches 1.
This behaviour is consistent with the trends reported in Ref. \citenum{DEVRIES2023106447}, where the influence of layer mass contrast on the LTC was identified at a phenomenological level.
Here, we provide a microscopic interpretation by linking this dependence to the layer-resolved character of the heat-carrying modes and its impact on relaxon transport properties.
It is instructive to contrast the preceding analysis against the average mass $\left<M\right>$ of each system.
Across the temperature range considered, heterobilayers with smaller $\left<M\right>$ tend to exhibit higher LTC, whereas when $\left<M\right>$ is larger the LTC is reduced (\autoref{fig:deltaP_combined}d).
However, the trend is not monotonic: systems with similar average mass can display significantly different conductivities, indicating that the average mass alone is not a sufficient descriptor to predict the LTC in isolation.
Taken together, these results suggest that thermal transport in heterobilayers is governed by an interplay between the overall mass scale and the layer mass contrast.
While lighter systems generally favour higher conductivity, a sufficiently large mass contrast is required to induce strong layer localisation of the relevant vibrational modes.
The layer mass contrast therefore provides a useful additional descriptor which, when combined with the average mass and the average phonon group velocity in the $[0,4]$ THz range, enables a physically motivated, a priori assessment of relative thermal conductivity across TMD heterobilayers.
\begin{figure*}[t]
\centering
\begin{minipage}{0.49\textwidth}
    \centering
    \includegraphics[width=\linewidth]{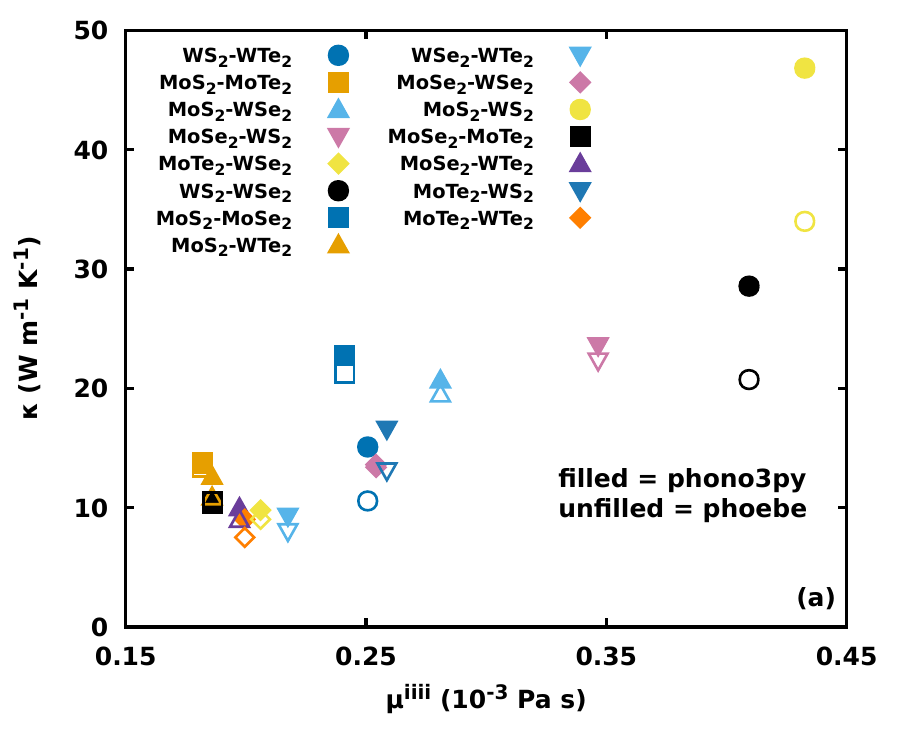}
\end{minipage}
\hfill
\begin{minipage}{0.49\textwidth}
    \centering
    \includegraphics[width=\linewidth]{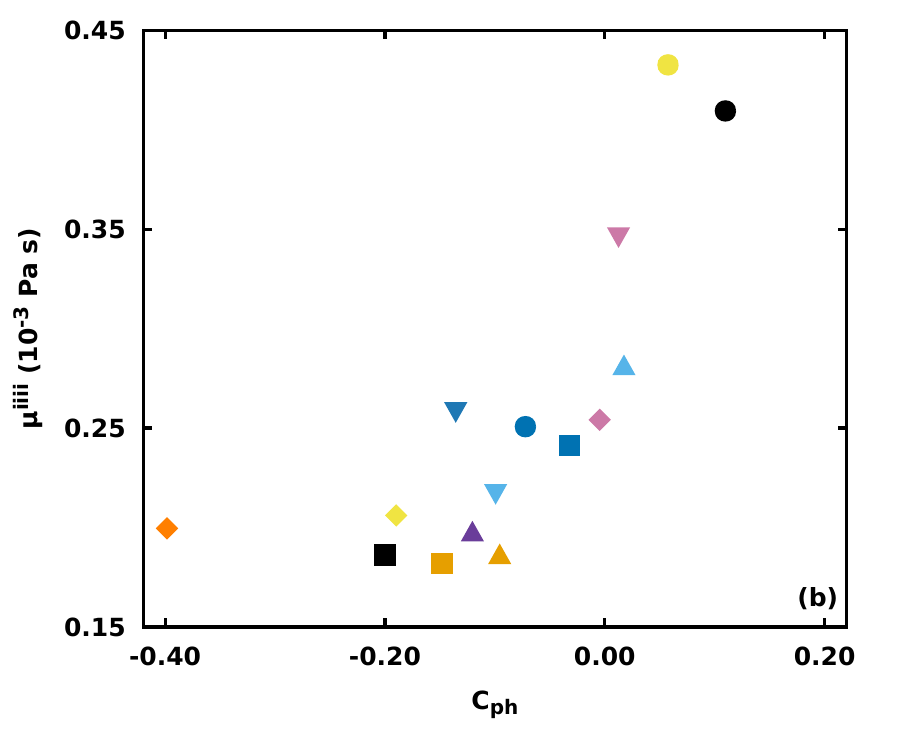}
\end{minipage}
\caption{
(a) Maximum (in-plane) conductivity, $\kappa = \kappa_{\text{max}} = \kappa_{\text{min}}$, as a function of the maximum component of thermal viscosity $\mu^{iiii}$.
(b) $\mu^{iiii}$ as a function of the M-X cophonicity $C_{\text{ph}}$, i.e. the relative overlap of the atom-projected phonon densities of states between M and X atoms.
}
\label{fig:viscosity}
\end{figure*}
So far, we have not considered the transport regime underlying the lattice thermal conductivity of the pristine TMD heterobilayers.
In general, thermal transport may be classified into ballistic, hydrodynamic, and diffusive regimes, depending on the relative importance of phonon scattering mechanisms.
Considering only intrinsic scattering, these are primarily Normal (momentum-conserving) and Umklapp (momentum-dissipating) processes, neglecting extrinsic contributions such as isotope, defect or boundary scattering.
The accessibility of each regime depends on both temperature and geometry.
In particular, phonon hydrodynamics --- characterised by the dominance of Normal processes --- has been shown to persist up to room temperature and beyond in two-dimensional materials such as graphene, hexagonal boron nitride (hBN), and molybdenum disulphide (MoS$_2$) \cite{Cepellotti2015}.
A similar dominance of Normal scattering over a broad temperature range from $[100,1000]$ K has been reported for monolayer MoS$_2$, WS$_2$, MoSe$_2$ and WSe$_2$ \cite{Torres_2019}.
Such classifications are typically established by analysing the relative contributions of of Normal and Umklapp scattering channels within the phonon BTE.
Now, it is also known that within the hydrodynamic regime, where momentum-conserving processes dominate, phonons exhibit collective behaviour characterised by a finite drift velocity.
In this context, the response of the crystal momentum flux to spatial gradients of the drift velocity defines a fourth-order thermal viscosity tensor, $\mu^{ijkl}$ \cite{PhysRevX.10.011019}.
Within the relaxon framework, thermal conductivity and thermal viscosity originate from the same underlying scattering operator but are associated with distinct parity sectors:
odd relaxons contribute to heat transport, whereas even relaxons determine the viscous response.
Consequently, thermal viscosity does not directly contribute to the heat current, but provides a probe of the degree of collective (hydrodynamic) transport encoded in the scattering operator.
Motivated by this, we consider the LTC as a function of the largest viscosity component $\mu^{iiii}$  which we find to be isotropic along the crystallographic zigzag and armchair directions ($\mu^{0000} = \mu^{1111}$).
As shown in \autoref{fig:viscosity}a, we observe an approximately linear correlation between the LTC and $\mu^{iiii}$ across the set of heterobilayers.
This suggests that, comparing among the heterobilayers at a particular temperature (here 300 K), an enhanced signature of collective transport correlates with increased LTC.
Similarly as for the LTC trends, we would like to understand whether simple descriptors of the phonon band structure can capture the observed behaviour.
In particular, we consider the cophonicity \cite{cophonicity}
\begin{equation}
 C_\text{ph} = \frac{\int_0^{4} \omega g^{(\text{M})}(\omega) \mathrm{d}\omega}{\int_0^{4} g^{(\text{M})}(\omega) \mathrm{d}\omega} - \frac{\int_0^{4} \omega g^{(\text{X})}(\omega) \mathrm{d}\omega}{\int_0^{4} g^{(\text{X})}(\omega) \mathrm{d}\omega},
\end{equation}
which quantifies the difference in the frequency-weighted centres of mass of the atom-projected phonon densities of states (DOS) for the metal ($g^{(M)}(\omega)$) and chalcogen ($g^{(X)}(\omega)$) sublattices over the $[0,4]$ THz frequency range.
A value close to zero indicates similar spectral distributions for both species, wheras positive (negative) values indicate that the metal (chalcogen) atoms dominate higher-frequency modes.
As seen in \autoref{fig:viscosity}b, we find that the thermal viscosity correlates approximately linearly with the cophonicity.
Specifically, systems in which the metal atoms contribute more strongly to higher-frequency modes exhibit larger thermal viscosity and, correspondingly, higher LTC.
This implies that the relative distribution of vibrational states between sublattices influences the balance between Normal and Umklapp scattering processes.
Overall, a larger thermal viscosity is indicative of a higher relative weight of momentum-conserving scattering processes, and therefore a stronger degree of collective transport.
We note that cophonicity relies only on harmonic phonon calculations, which are less computational demanding than evaluating explicitly thermal conductivity and viscosity;
as a consequence, cophonicity can then be used as a descriptor for a preliminary identification of promising candidate TMD van der Waals heterostructures with target (e.g. high) thermal conductivity and viscosity in rapid high-throughput screenings of databases.
Since the relative importance of Normal and Umklapp processes also governs the sensitivity of thermal transport to boundary and defect scattering \cite{10.1088/978-0-7503-1738-2ch1}, this analysis provides a physically motivated basis for anticipating how transport behaviour may evolve with system size and disorder.
\subsection{Doped heterobilayers}
\label{sec:doped_systems}
\begin{figure*}[t]
\centering
\begin{minipage}{0.49\textwidth}
    \centering
    \includegraphics[width=\linewidth]{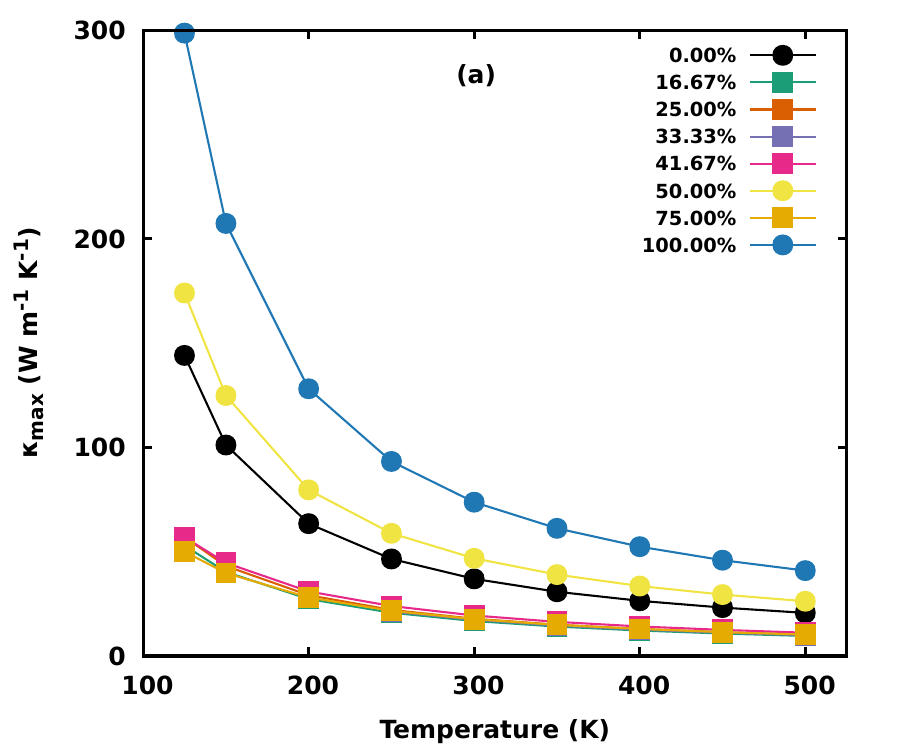}
\end{minipage}
\hfill
\begin{minipage}{0.49\textwidth}
    \centering
    \includegraphics[width=\linewidth]{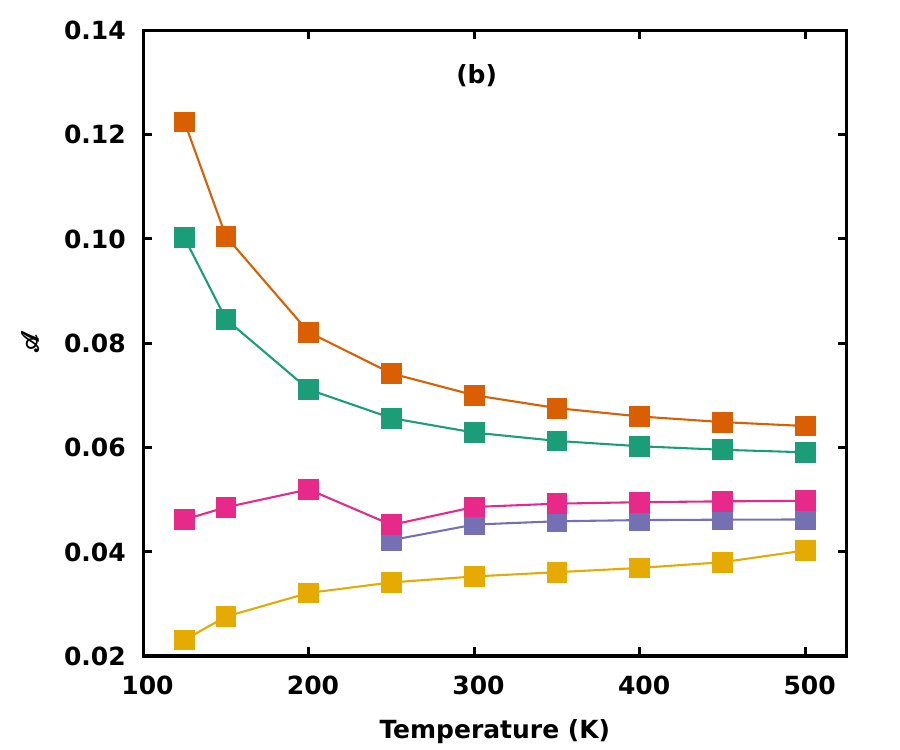}
\end{minipage}
\caption{Thermal transport properties of doped heterobilayers bounded by MoS$_2$ and WS$_2$ homobilayers, with W treated as the dopant.
(a) Maximum in-plane conductivity, $\kappa_{\text{max}}$, as a function of temperature.
(b) Anisotropy, $\mathcal{A}$, between the maximum ($\bm{e}_{\text{max}}$) and minimum ($\bm{e}_{\text{min}}$) transport directions.
}
\label{fig:ltc_vs_dop_combined}
\end{figure*}

\begin{figure}[t]
\centering
\includegraphics[width=\columnwidth]{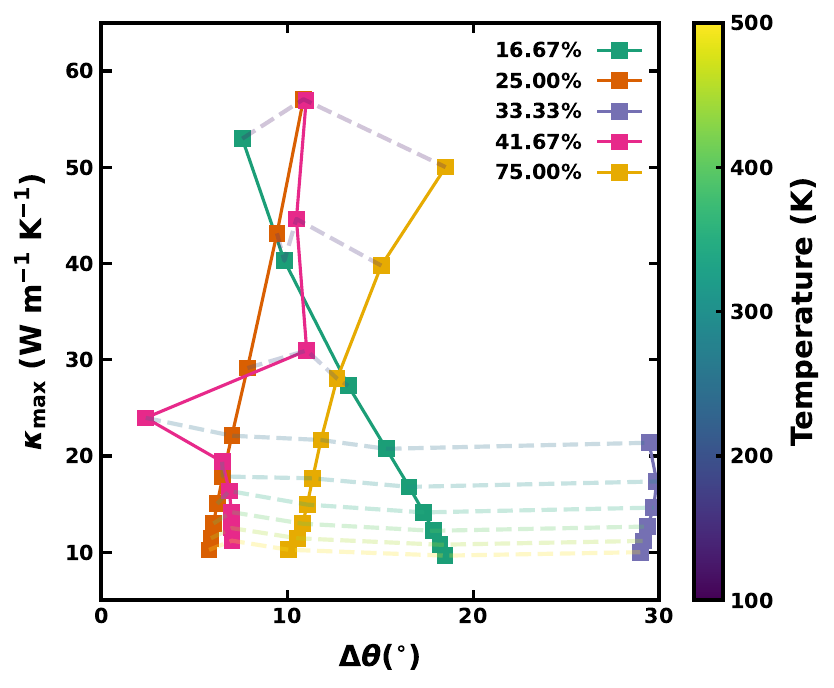}
\caption{
Maximum (in-plane) LTC $\kappa_{\text{max}}$ plotted against $\Delta \theta$, which describes the deviation of the direction of maximum LTC $\bm{e}_{\text{max}}$ from the nearest zigzag direction.
Data-points from different systems at the same temperatures are connected by a faint dashed coloured line, with temperature set according to the colour bar.
}
\label{fig:ltc_vs_dtheta}
\end{figure}

We now shift our focus onto W-doped Mo$_{1-x}$W$_{x}$S$_2$ TMD heterobilayers, for a selection of configurations with $x \in [0.0,1.0]$.
Recall that $x$ denotes the W content on a fractional scale, such that $x = 0.0$ and $x = 1.0$ correspond to the homobilayer MoS$_2$ and WS$_2$ respectively, while $x = 0.5$ corresponds to the Mos$_2$-WS$_2$  heterobilayer.
Herein we will refer to dopant concentration using their percentage equivalent value $100x$\%.
We begin by repoting the maximum LTC as a function of temperature (\autoref{fig:ltc_vs_dop_combined}a), obtained from \textsc{phono3py} LBTE calculations.
The temperature range considered is $[125,500]$ K, as opposed to $[100,500]$ K used for the pristine heterobilayers, as at $100$ K we did not achieve convergence of the LTC.
At 33\% concentration of dopant, convergence of the LTC is found only at $250$ K and above.
We find that thermal transport in the homobilayers is isotropic in-plane, similarly as for heterobilayer MoS$_2$-WS$_2$.
Further, the LTC of the heterobilayer lies between the homobilayers, which is an expected outcome based on previous results in the literature \cite{C8RA10601K}.
Conversely, for the doped systems we find in-plane thermal transport to be anisotropic, that is $\kappa_{\text{max}} \neq \kappa_{\text{min}}$.
Looking at $\kappa_{\text{max}}$ we observe a reduction in the maximum LTC for all doped systems across the whole temperature range in comparison to heterobilayer MoS$_2$-WS$_2$ and the homobilayers.
Further, there does not appear to a trivial correlation between dopant concentration and the LTC; we see that the values are very similar at each temperature for all doped systems.
Notably, the deviation between the doped systems and the homobilayers/heterobilayers grows at lower temperatures.
This can be attributed to the suppression of Umklapp scattering in this regime, such that mass-disorder scattering becomes a dominant resistive mechanism in the doped systems \cite{PhysRevB.88.144306}, limiting the otherwise long phonon lifetimes and reducing the thermal conductivity.
Then, to quantify the anisotropy we use the normalised ratio
\begin{equation}
 \mathcal{A} = \frac{\kappa_{\text{max}}-\kappa_{\text{min}}}{\kappa_{\text{max}}+\kappa_{\text{min}}}
\end{equation}
such that $\mathcal{A} \rightarrow 0$ describes a perfectly isotropic system, and $\mathcal{A} \rightarrow 1$ describes a strongly anisotropic system where $\kappa_{\text{max}} \gg \kappa_{\text{min}}$.
We find (\autoref{fig:ltc_vs_dop_combined}b) that $\mathcal{A}$ varies most strongly at low temperature and is system dependent.
For example, at $25.00$\% concentration of W the anisotropy reaches a maximum value of $\mathcal{A} \approx 0.12$ at $T=125$ K; instead, at $75.00$\% concentration the anisotropy is minimised at the same temperature ($\mathcal{A} \approx 0.02$).
As temperature rises, $\mathcal{A}$ increases at $41.67$\% and $75.00$\% concentration, meanwhile at $16.67$\%, $25.00$\% the anisotropy decreases; this suggests that $\mathcal{A}$ stabilises at higher temperature.
A similar stability is observed at 33\% concentration beyond 250 K.
Considering all doped systems, the scale of the variation over the whole temperature range is not large, at a maximum value of $\Delta \mathcal{A} \approx 0.05$ for W at $25$\%.
This trend is consistent with the increased importance of mass-disorder scattering at low temperatures, where intrinsic anharmonic scattering is suppressed.
Although one might infer a concentration-dependent inversion in the temperature dependence of the anisotropy, this interpretation is complicated by the fact that the dopant distribution varies between configurations.
We expect that this plays an important role in determining how the anisotropy evolves with temperature, which deserves further attention.
This factor was not studied in detail here, since we focus on a physically-motivated selection of dopant concentrations and distributions which were determined to have negative formation energy with respect to homobilayer MoS$_2$ and WS$_2$ \cite{PhysRevMaterials.8.106001}.
Recalling from \autoref{sec:methods}, for anisotropic systems we define a frame-invariant reference angle $\Delta \theta \in [0,30^\circ]$ as the smallest angle between the direction of maximum conductivity $\bm{e}_{\text{max}}$ and the nearest zigzag direction.
In the doped systems, $\Delta \theta$ varies with temperature, indicating a rotation of the principal axes of thermal transport.
This reflects the breaking of in-plane symmetry induced by the dopant distribution, combined with the temperature dependence of phonon populations and scattering processes, which modifies the relative contributions to thermal transport along different crystallographic directions.
As shown in \autoref{fig:ltc_vs_dtheta}, the relationship between $\kappa_{\text{max}}$ and $\Delta \theta$ depends strongly on dopant concentration.
For example, at $25.00$\% and $75.00$\% concentration, $\kappa_{\text{max}}$ increases approximately linearly with $\Delta \theta$, whereas at $16.67$\% the opposite is true.
At $41.67$\% concentration the dependence is weak at higher temperatures, and we observe strong variation in $\Delta \theta$ at lower temperatures, despite the maximum LTC varying monotonically with temperature.
Although we don't have low temperature data-points at 33\% concentration, it initially appears stable at a $30^{\circ}$ rotation, indicating that heat prefers to flow preferentially along one armchair direction.
We don't specify the specific armchair direction, since $\Delta\theta$ is defined within the symmetry-restricted interval $[0,30^\circ]$, such that when the principal direction ($\bm{e}_{\text{max}}$) crosses between symmetry-equivalent sectors the angle is folded back.
This description is preferred since it provides a symmetry-invariant measure of the transport direction relative to the crystallographic axes, whereas the absolute angle does not directly reflect this equivalence.
In general, these results suggest a route to tuning the orientation of maximum heat conduction via composition and temperature.
However, the magnitude of the anisotropy remains modest across the dopant concentration range considered, such that directional control is limited by the relatively small contrast between $\kappa_{\text{max}}$ and $\kappa_{\text{min}}$.
Further, a key limitation in translating these results to experiment is that the temperature dependence of $\Delta \theta$ cannot be predicted from dopant concentration alone. 
In the present calculations, each concentration corresponds to a distinct dopant configuration, such that variations in $\Delta \theta$ also reflect differences in the spatial arrangement of dopants, rather than concentration alone. 
The configurations considered here correspond to low-energy structures at 0 K, whereas experimental realisations are likely to exhibit disorder and variability in atomic arrangement.
A systematic investigation of configurational disorder at fixed composition is therefore required to establish the robustness of these directional trends.
Now, rather than performing further relaxon BTE calculations, we assess the extent to which the descriptors identified in the previous section ({\autoref{sec:prist_hetbi}) remain predictive in the presence of disorder due to doping.
We provide the high-symmetry path band structures for reference in the Supporting Information section S2, since they are not integral to the discussion here.
First, we examine the root mean square average phonon group velocities projected onto the direction of maximum transport, $v_{\text{rms}} = \sqrt{\left\langle(\bm{v}_g \cdot \bm{e}_{\text{max}})^2\right\rangle}$, and plot $\kappa_{\text{max}}$ as a function of $v_{\text{rms}}$ in the $[0,4]$ THz frequency range (\autoref{fig:dop_mode_localisation}a).
In comparison to pristine heterobilayer MoS$_2$-WS$_2$, we observe a clear reduction in $v_{\text{rms}}$.
This reflects the modification of the phonon dispersion induced by mass disorder and symmetry breaking, and contributes directly to the reduction of the LTC.
In addition to this reduction in group velocity, the presence of mass disorder introduces additional resistive scattering channels.
Although phonon lifetimes are not explicitly analysed here, they are expected to be suppressed by this mechanism, which is a further contributor to the reduction in thermal transport.
Second, we study the dependence of the maximum LTC on the layer localisation of each phonon mode $\Delta P_\lambda$ (recall \autoref{eq:layer_localisation}), through its transport-weighted (phonon) modal average $\left<|\Delta P_\lambda|\right>_\kappa$ (calculated via \autoref{eq:transport_weighting}).
As shown in \autoref{fig:dop_mode_localisation}b, we observe no clear correlation between $\kappa_{\text{max}}$ and $\left<|\Delta P_\lambda|\right>_\kappa$, despite the latter varying systematically with the layer mass contrast $|\Delta M|$ (\autoref{fig:dop_mode_localisation}c).
This contrasts with the trends observed for the pristine heterobilayers (\autoref{fig:deltaP_combined}a), where $\left<|\Delta P_\lambda|\right>_\kappa$ was found to correlate with the LTC.
Nevertheless, the magnitude of $\left<|\Delta P_\lambda|\right>_\kappa$ remains comparable across the doped systems and heterobilayer MoS$_2$-WS$_2$, indicating that phonon modes retain a well-defined layer character despite the presence of mass disorder.
Finally, we find no strong depence between the LTC and the cophonicity $C_{\text{ph}}$ for the doped systems; it is worth noting in passing that the homobilayers and MoS$_2$-WS$_2$ heterobilayer exhibit a positive correlation here, which suggests that $C_{\text{ph}}$ has predictive value for the LTC beyond the inter-system heterobilayer comparisons of \autoref{sec:prist_hetbi}.
Returning to the doped heterobilayers, the contribution of a given phonon mode to thermal transport is no longer determined primarily by its spatial character, but instead by its interactions with other modes through scattering processes.
The absence of correlation with the maximum LTC therefore does not arise from a loss of localisation, but rather from a decoupling between mode character and thermal transport.
A more complete description would therefore require analysis in the relaxon basis, as implemented in \textsc{phoebe}, which provides direct access to scattering-driven transport properties.
Regardless, the present analysis demonstrates that the reduction in thermal conductivity can be qualitatively rationalised in terms of reduced group velocities and enhanced scattering within a phonon-based framework.
\begin{figure*}[t]
\centering

\begin{minipage}{0.49\textwidth}
    \centering
    \includegraphics[width=\linewidth]{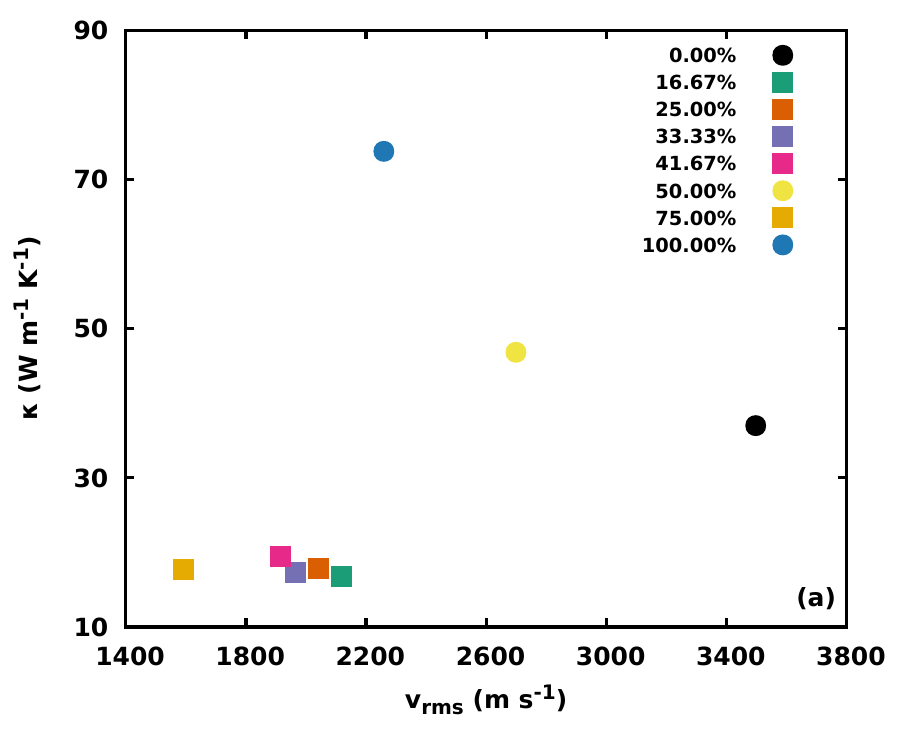}
\end{minipage}
\hfill
\begin{minipage}{0.49\textwidth}
    \centering
    \includegraphics[width=\linewidth]{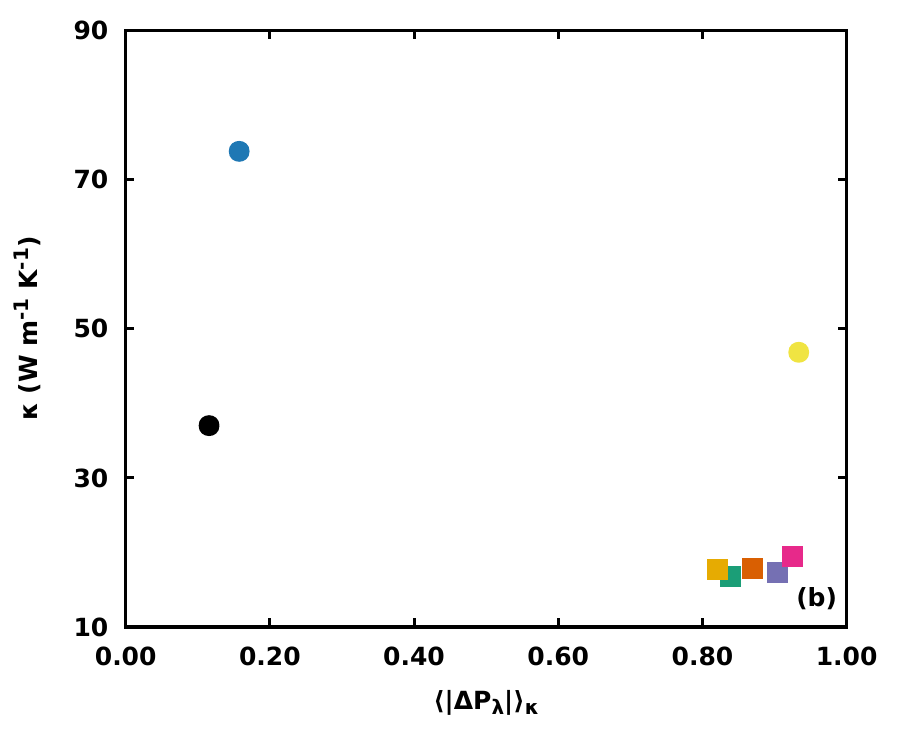}
\end{minipage}
\vspace{0.5em}
\begin{minipage}{0.49\textwidth}
    \centering
    \includegraphics[width=\linewidth]{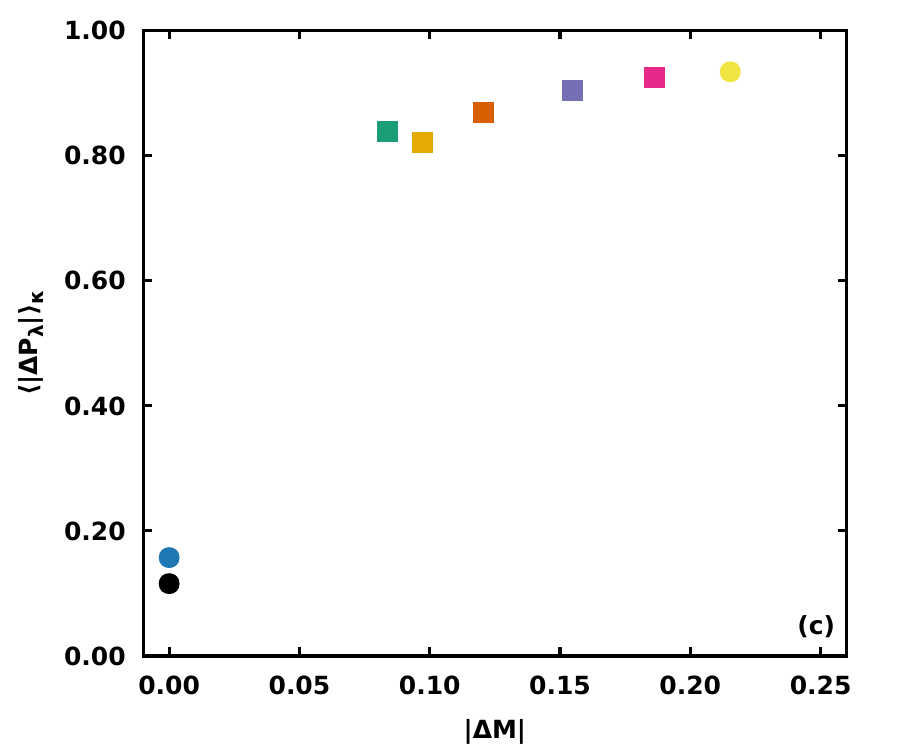}
\end{minipage}
\hfill
\begin{minipage}{0.49\textwidth}
    \centering
    \includegraphics[width=\linewidth]{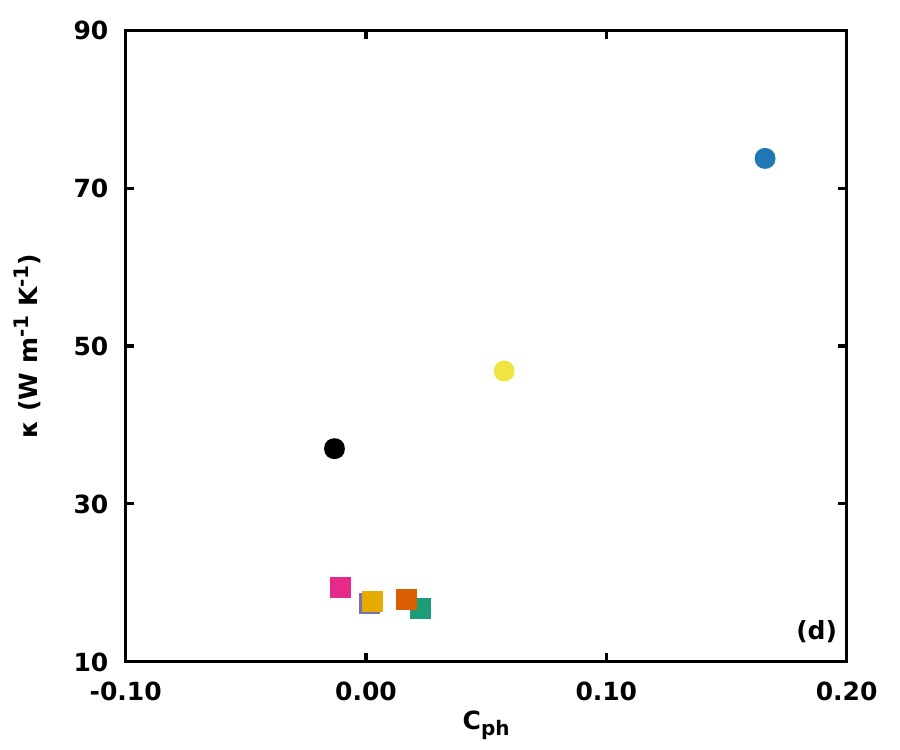}
\end{minipage}
\caption{
(a) Maximum in-plane conductivity $\kappa_{\mathrm{max}}$ as a function of the average phonon group velocity projected onto the direction of maximum transport, $v_{\text{rms}}$, evaluated over the $[0,4]\,\mathrm{THz}$ frequency range.
(b) $\kappa_{\mathrm{max}}$ as a function of the transport-weighted modal average layer mode localisation $\left\langle |\Delta P_\lambda| \right\rangle_\kappa$.
(c) $\left\langle |\Delta P_\lambda| \right\rangle_\kappa$ as a function of the layer mass contrast $|\Delta M|$.
(d) $\kappa_{\mathrm{max}}$ plotted against the M-X cophonicity $C_{\text{ph}}$.
}
\label{fig:dop_mode_localisation}
\end{figure*}
\section{Conclusions}
\label{sec:conclusions}

In this work, we investigate the lattice thermal conductivity (LTC) of TMD heterobilayers of the form  MX$_2$-M$^\prime$X$_2^\prime$, where M,M$^\prime \in \{\text{Mo, W}\}$ and X,X$^\prime \in \{\text{S, Se, Te}\}$.
Excluding homobilayers and treating layer order as equivalent, we study 15 unique heterobilayers;
in addition, we consider W-doped Mo$_{1-x}$W$_{x}$S$_2$.
For the heterobilayers, we calculate the LTC via exact solution of the LBTE at varying temperature in a phonon basis including first-order anharmonic scattering, and observe isotropic in-plane transport and a preserved ordering of pristine heterobilayers across the considered temperature range.
We then calculate LTC at $300$ K via exact solution of the LBTE in a relaxon basis.
Through analysis of the relaxon solution, we relate the underlying transport physics to descriptors based on harmonic phonons, including the root mean square average phonon group velocity and the phonon mode layer localisation.
We use the thermal viscosity as a probe of the degree of collective (hydrodynamic) transport encoded in the scattering operator, and show via the cophonicity that the relative distribution of vibrational states between sublattices influences the balance between Normal and Umklapp scattering processes.
Cophonicity can then be used as a low-computationally-demanding descriptor for rapid screening of databases to identify candidate TMD heterostructures with target thermal conductivity and viscosity.

We then assess the transferability of the above descriptors to the doped heterobilayers, performing LTC calculations at varying temperature.
While the phonon group velocities remain indicative of LTC trends, the layer localisation and cophonicity no longer correlate in general with the thermal conductivity, reflecting the increased importance of intra-layer scattering due to mass disorder after inclusion of the dopant.
We further find that thermal transport in the doped systems is weakly anisotropic, and that the direction of maximum conductivity varies with temperature and dopant concentration. 
This is quantified using the symmetry-invariant angle $\Delta \theta$.
This suggests a possible route to tuning the direction of maximum (and minimum) thermal transport via concentration and temperature. 
However, the applicability is limited by the modest magnitude of the anisotropy and by the fact that each concentration corresponds to a distinct dopant configuration, such that variations in $\Delta \theta$ also reflect differences in the spatial arrangement of dopants.
In realistic systems, configurational disorder and variability in atomic arrangement are expected.
A systematic investigation of configurational disorder at fixed composition is therefore required to establish the robustness of the predicted directional trends and to assess the extent to which anisotropy can be enhanced for practical applications.
%
\begin{acknowledgments}

This work was co-funded by the European Union under the project ``Robotics and advanced industrial production'' (reg. no. CZ.02.01.01/00/22\_008/0004590), by the Czech Science Foundation project No. 23-07785S, and by the Ministry of Education, Youth and Sports of the Czech Republic through the e-INFRA CZ (ID: 90254).
The use of the \textsc{vesta} software is also acknowledged for generating crystal structure visualisations \cite{VESTA}.

\end{acknowledgments}


%




\clearpage
\onecolumngrid
\begin{center}
{\Large \bfseries Supporting Information}
\end{center}

\vspace{1em}

\setcounter{figure}{0}

\renewcommand{\thefigure}{S\arabic{figure}}

\setcounter{section}{0}
\renewcommand{\thesection}{S\arabic{section}}
\renewcommand{\thesubsection}{S\arabic{section}.\arabic{subsection}}
\renewcommand{\thesubsubsection}{S\arabic{section}.\arabic{subsection}.\arabic{subsubsection}}


\section{Optimised geometries}
\label{sec:structopt}
We report the optimised geometries of transition metal dichalcogenide (TMD) pristine heterobilayers in section S1.1, pristine homobilayers in section S1.2, and doped heterobilayers in section S1.3 calculated using the Vienna Ab initio Simulation Package (\textsc{vasp}).
Lattice parameters are provided in Angstrom (\AA{}), and atomic positions in fractional coordinates.

\subsection{Pristine heterobilayers}
\label{sec:heterobilayers}

\subsubsection{MoSe$_2$-MoTe$_2$}
\label{struct:msemte}
\begin{center}
\begin{verbatim}
MoSe2-MoTe2
   1.00000000000000     
     3.4165023429852939    0.0000000000000005    0.0000000000000000
    -1.7082511714926469    2.9587778211143205    0.0000000000000000
     0.0000000000000000    0.0000000000000000   40.0000000000000071
  Mo              Se              Te            
     2     2     2
Direct
     0.6666666666666643    0.3333333333333357    0.9094554335438042
     0.3333333333333357    0.6666666666666643    0.7402463870984808
     0.3333333333333357    0.6666666666666643    0.8685415491453337
     0.3333333333333357    0.6666666666666643    0.9503518484731466
     0.6666666666666643    0.3333333333333357    0.6938091340515391
     0.6666666666666643    0.3333333333333357    0.7867080063376942
 
\end{verbatim}
\end{center}

\subsubsection{MoSe$_2$-WSe$_2$}
\label{struct:msewse}
\begin{center}
\begin{verbatim}
MoSe2-WSe2
   1.00000000000000     
     3.3268390942687827   -0.0000000000000002    0.0000000000000000
    -1.6634195471343922    2.8811271699399761    0.0000000000000000
     0.0000000000000000    0.0000000000000000   40.0000000000000071
  W               Se              Mo            
     1     4     1
Direct
     0.6666666666666643    0.3333333333333357    0.7550818205261686
     0.6666666666666643    0.3333333333333357    0.9585872486968603
     0.0000000000000000    0.0000000000000000    0.7132464692474060
     0.0000000000000000    0.0000000000000000    0.7968496839300502
     0.6666666666666643    0.3333333333333357    0.8753409972497153
     0.0000000000000000    0.0000000000000000    0.9169544247247957
 
\end{verbatim}
\end{center}

\subsubsection{MoSe$_2$-WS$_2$}
\label{struct:msews}
\begin{center}
\begin{verbatim}
MoSe2-WS2
   1.00000000000000     
     3.2540268732970983   -0.0000000000000009    0.0000000000000000
    -1.6270134366485489    2.8180699368725333    0.0000000000000000
     0.0000000000000000    0.0000000000000000   40.0000000000000071
  Mo              Se              W               S             
     1     2     1     2
Direct
     0.3333333333333357    0.6666666666666643    0.0843498354129645
     0.6666666666666643    0.3333333333333357    0.1266040058720063
     0.6666666666666643    0.3333333333333357    0.0420778976130506
     0.6666666666666643    0.3333333333333357    0.2420734765537731
     0.3333333333333357    0.6666666666666643    0.2035440771820059
     0.3333333333333357    0.6666666666666643    0.2806961183911936
 
\end{verbatim}
\end{center}

\subsubsection{MoSe$_2$-WTe$_2$}
\label{struct:msewte}
\begin{center}
\begin{verbatim}
MoSe2-WTe2
   1.00000000000000     
     3.4340081510849467   -0.0000000000000010    0.0000000000000000
    -1.7170040755424743    2.9739382956423959    0.0000000000000000
     0.0000000000000000    0.0000000000000000   40.0000000000000071
  W               Te              Mo              Se            
     1     2     1     2
Direct
     0.3333333333333357    0.6666666666666643    0.2644926837536470
     0.0000000000000000    0.0000000000000000    0.2180700863001319
     0.0000000000000000    0.0000000000000000    0.3109406257739947
     0.0000000000000000    0.0000000000000000    0.0964988510292486
     0.3333333333333357    0.6666666666666643    0.1372637777805747
     0.3333333333333357    0.6666666666666643    0.0557396003373974
 
\end{verbatim}
\end{center}

\subsubsection{MoS$_2$-MoSe$_2$}
\label{struct:msmse}
\begin{center}
\begin{verbatim}
MoS2-MoSe2
   1.00000000000000     
     3.2525041784888455    0.0000000000000009    0.0000000000000000
    -1.6262520892444228    2.8167512444863751    0.0000000000000000
     0.0000000000000000    0.0000000000000000   40.0000000000000071
  Mo              Se              S             
     2     2     2
Direct
     0.3333333333333357    0.6666666666666643    0.0836321930985852
     0.6666666666666643    0.3333333333333357    0.2428196219351455
     0.3333333333333357    0.6666666666666643    0.2005371186182391
     0.3333333333333357    0.6666666666666643    0.2851097733588461
     0.6666666666666643    0.3333333333333357    0.1220617203659969
     0.6666666666666643    0.3333333333333357    0.0451849836481810
 
\end{verbatim}
\end{center}

\subsubsection{MoS$_2$-MoTe$_2$}
\label{struct:msmte}
\begin{center}
\begin{verbatim}
MoS2-MoTe2
   1.00000000000000     
     3.3454689999956151   -0.0000000000000000    0.0000000000000000
    -1.6727344999978075    2.8972611415695275    0.0000000000000000
     0.0000000000000000    0.0000000000000000   40.0000000000000071
  Mo              Te              S             
     2     2     2
Direct
     0.6666666666666643    0.3333333333333357    0.9073712340668803
     0.3333333333333357    0.6666666666666643    0.7423246542727675
     0.6666666666666643    0.3333333333333357    0.7895182184811146
     0.6666666666666643    0.3333333333333357    0.6951765789709642
     0.3333333333333357    0.6666666666666643    0.8695889836676562
     0.3333333333333357    0.6666666666666643    0.9451326891906225
 
\end{verbatim}
\end{center}

\subsubsection{MoS$_2$-WSe$_2$}
\label{struct:mswse}
\begin{center}
\begin{verbatim}
MoS2-WSe2
   1.00000000000000     
     3.2557680457380651    0.0000000000000011    0.0000000000000000
    -1.6278840228690319    2.8195778364387802    0.0000000000000000
     0.0000000000000000    0.0000000000000000   40.0000000000000071
  W               Se              Mo              S             
     1     2     1     2
Direct
     0.6666666666666643    0.3333333333333357    0.7569959370878605
     0.0000000000000000    0.0000000000000000    0.7145042508872660
     1.0000000000000000    0.0000000000000000    0.7994350514336978
     0.0000000000000000    0.0000000000000000    0.9150314840323884
     0.6666666666666643    0.3333333333333357    0.9534528960598907
     0.6666666666666643    0.3333333333333357    0.8766410248738931
 
\end{verbatim}
\end{center}

\subsubsection{MoS$_2$-WS$_2$}
\label{struct:msws}
\begin{center}
\begin{verbatim}
MoS2-WS2
   1.00000000000000     
     3.1889225289023031   -0.0000000000000012    0.0000000000000000
    -1.5944612644511518    2.7616879207299125    0.0000000000000000
     0.0000000000000000    0.0000000000000000   39.9862299999999991
  Mo              S               W             
     1     4     1
Direct
     0.0000000000000000    0.0000000000000000    0.2322228344817223
     0.6666666666666643    0.3333333333333357    0.1933001501979815
     0.0000000000000000    0.0000000000000000    0.1172160270439588
     0.0000000000000000    0.0000000000000000    0.0389270938066464
     0.6666666666666643    0.3333333333333357    0.2711830259970060
     0.6666666666666643    0.3333333333333357    0.0781108684776932
 
\end{verbatim}
\end{center}

\subsubsection{MoS$_2$-WTe$_2$}
\label{struct:mswte}
\begin{center}
\begin{verbatim}
MoS2-WTe2
   1.00000000000000     
     3.3581232259427378    0.0000000000000001    0.0000000000000000
    -1.6790616129713689    2.9082200227049548    0.0000000000000000
     0.0000000000000000    0.0000000000000000   40.0000000000000071
  W               Te              Mo              S             
     1     2     1     2
Direct
     0.3333333333333357    0.6666666666666643    0.2624665517841697
     0.0000000000000000    0.0000000000000000    0.2152362524940919
     0.0000000000000000    0.0000000000000000    0.3096919203235951
     0.0000000000000000    0.0000000000000000    0.0985317840563491
     0.3333333333333357    0.6666666666666643    0.1362218126726641
     0.3333333333333357    0.6666666666666643    0.0608573036441242
 
\end{verbatim}
\end{center}

\subsubsection{MoTe$_2$-WSe$_2$}
\label{struct:mtewse}
\begin{center}
\begin{verbatim}
MoTe2-WSe2
   1.00000000000000     
     3.4301487100804993    0.0000000000000008    0.0000000000000000
    -1.7150743550402496    2.9705959216881364    0.0000000000000000
     0.0000000000000000    0.0000000000000000   40.0000000000000071
  W               Se              Mo              Te            
     1     2     1     2
Direct
     0.6666666666666643    0.3333333333333357    0.7520696978486827
     0.0000000000000000    0.0000000000000000    0.7111276661220755
     0.0000000000000000    0.0000000000000000    0.7929400874472564
     0.0000000000000000    0.0000000000000000    0.9199784513015187
     0.6666666666666643    0.3333333333333357    0.9662771659989428
     0.6666666666666643    0.3333333333333357    0.8736675756565189
 
\end{verbatim}
\end{center}

\subsubsection{MoTe$_2$-WS$_2$}
\label{struct:mtews}
\begin{center}
\begin{verbatim}
MoTe2-WS2
   1.00000000000000     
     3.3524986353322310   -0.0000000000000003    0.0000000000000000
    -1.6762493176661160    2.9033489843503753    0.0000000000000000
     0.0000000000000000    0.0000000000000000   40.0000000000000071
  Mo              Te              W               S             
     1     2     1     2
Direct
     0.6666666666666643    0.3333333333333357    0.9066962509914478
     0.3333333333333357    0.6666666666666643    0.8595998059635972
     0.3333333333333357    0.6666666666666643    0.9537597369216345
     0.3333333333333357    0.6666666666666643    0.7430489120027675
     0.6666666666666643    0.3333333333333357    0.7808134329481426
     0.6666666666666643    0.3333333333333357    0.7051942198224157
 
\end{verbatim}
\end{center}

\subsubsection{MoTe$_2$-WTe$_2$}
\label{struct:mtewte}
\begin{center}
\begin{verbatim}
MoTe2-WTe2
   1.00000000000000     
     3.5443595003731061    0.0000000000000009    0.0000000000000000
    -1.7721797501865526    3.0695053674678325    0.0000000000000000
     0.0000000000000000    0.0000000000000000   40.0000000000000071
  W               Te              Mo            
     1     4     1
Direct
     0.3333333333333357    0.6666666666666643    0.2674810069227063
     0.0000000000000000    0.0000000000000000    0.2221553472953998
     0.3333333333333357    0.6666666666666643    0.1387467061308457
     0.3333333333333357    0.6666666666666643    0.0482801675009910
     0.0000000000000000    0.0000000000000000    0.3128379394090412
     0.0000000000000000    0.0000000000000000    0.0935044577160104
 
\end{verbatim}
\end{center}

\subsubsection{WSe$_2$-WTe$_2$}
\label{struct:wsewte}
\begin{center}
\begin{verbatim}
WSe2-WTe2
   1.00000000000000     
     3.4483974240389403   -0.0000000000000012    0.0000000000000000
    -1.7241987120194706    2.9863997715625401    0.0000000000000000
     0.0000000000000000    0.0000000000000000   40.0000000000000071
  W               Te              Se            
     2     2     2
Direct
     0.6666666666666643    0.3333333333333357    0.0972519087531323
     0.0000000000000000    0.0000000000000000    0.2637663853129402
     0.6666666666666643    0.3333333333333357    0.2175175932801532
     0.6666666666666643    0.3333333333333357    0.3100644966228496
     0.0000000000000000    0.0000000000000000    0.1379463030441007
     0.0000000000000000    0.0000000000000000    0.0564589379618183
 
\end{verbatim}
\end{center}

\subsubsection{WS$_2$-WSe$_2$}
\label{struct:wswse}
\begin{center}
\begin{verbatim}
WS2-WSe2
   1.00000000000000     
     3.2575152190576300    0.0000000000000005    0.0000000000000000
    -1.6287576095288163    2.8210909329183389    0.0000000000000000
     0.0000000000000000    0.0000000000000000   40.0000000000000071
  W               Se              S             
     2     2     2
Direct
     0.3333333333333357    0.6666666666666643    0.9141651464372270
     1.0000000000000000    0.0000000000000000    0.7578397920773389
     0.3333333333333357    0.6666666666666643    0.7153684061915986
     0.3333333333333357    0.6666666666666643    0.8002334333841307
     0.0000000000000000    0.0000000000000000    0.9527602490525661
     0.0000000000000000    0.0000000000000000    0.8756936172321341
 
\end{verbatim}
\end{center}

\subsubsection{WS$_2$-WTe$_2$}
\label{struct:wswte}
\begin{center}
\begin{verbatim}
WS2-WTe2
   1.00000000000000     
     3.3662958986378064    0.0000000000000012    0.0000000000000000
    -1.6831479493189032    2.9152977648757026    0.0000000000000000
     0.0000000000000000    0.0000000000000000   40.0000000000000071
  W               Te              S             
     2     2     2
Direct
     0.6666666666666643    0.3333333333333357    0.0993631778305939
     0.0000000000000000    0.0000000000000000    0.2616711806650916
     0.6666666666666643    0.3333333333333357    0.2145609168295023
     0.6666666666666643    0.3333333333333357    0.3087997297234817
     0.0000000000000000    0.0000000000000000    0.1370019840905538
     0.0000000000000000    0.0000000000000000    0.0616086358357710
 
\end{verbatim}
\end{center}

\subsection{Pristine homobilayers}
\label{sec:homobilayers}

\subsubsection{MoS$_2$-MoS$_2$}
\label{struct:mos2}
\begin{center}
\begin{verbatim}
MoS2-MoS2
   1.00000000000000     
     3.1904242934985794   -0.0000000000000000    0.0000000000000000
    -1.5952121467492897    2.7629884870207890    0.0000000000000000
     0.0000000000000000    0.0000000000000000   39.9862300000000062
  Mo              S             
     2     4
Direct
     0.3333333333333357    0.6666666666666643    0.9222053823097890
     0.6666666666666643    0.3333333333333357    0.0777946176902111
     0.3333333333333357    0.6666666666666643    0.0388687207088780
     0.6666666666666643    0.3333333333333357    0.9611312792911221
     0.6666666666666643    0.3333333333333357    0.8832555272298448
     0.3333333333333357    0.6666666666666643    0.1167444727701551
 
\end{verbatim}
\end{center}

\subsubsection{WS$_2$-WS$_2$}
\label{struct:ws2}
\begin{center}
\begin{verbatim}
WS2-WS2
   1.00000000000000     
     3.1872375230084371   -0.0000000000000006    0.0000000000000000
    -1.5936187615042181    2.7602286628202952    0.0000000000000000
     0.0000000000000000    0.0000000000000000   40.1103900000000024
  S               W             
     4     2
Direct
     0.3333333333333357    0.6666666666666643    0.0369956985189263
     0.6666666666666643    0.3333333333333357    0.9630043014810737
     0.6666666666666643    0.3333333333333357    0.8849604382811078
     0.3333333333333357    0.6666666666666643    0.1150395617188921
     0.3333333333333357    0.6666666666666643    0.9240371611197656
     0.6666666666666643    0.3333333333333357    0.0759628388802344
 
\end{verbatim}
\end{center}

\subsection{Doped heterobilayers}
\label{sec:doped}

\subsubsection{16.67\%}
\label{struct:16}
\begin{center}
\begin{verbatim}
Mo{1-x}S2-W{x}S2 @ 16.67% concentration of W
   1.00000000000000     
     6.3831677672722069    0.0044534052785647    0.0000000000000000
    -1.6005231177776971    8.2902042874049844    0.0000000000000000
     0.0000000000000024    0.0000000000000030   39.9862299999999991
  Mo              S               W             
    10    24     2
Direct
     0.6110727635405321    0.4444358591572197    0.2326848636998363
     0.7777407578912022    0.1111171808918051    0.2326837874600780
     0.8901115624139443    0.5569422288203867    0.0776139804453762
     0.1110956394387374    0.4444504185843147    0.2326907642245334
     0.0553962679566150    0.2223854641820964    0.0776089607359348
     0.2777631528353156    0.1111074136345895    0.2326894034726807
     0.2212598483617975    0.8885485465761599    0.0776015556456124
     0.4444161566943025    0.7777541278040226    0.2326966360342123
     0.7231205861450178    0.8886244496409961    0.0775988681600483
     0.9444187382589796    0.7777811205749113    0.2326967882656060
     0.3888498066195790    0.5555467319847224    0.1937648778076157
     0.6114077961185601    0.4444443808798095    0.1166152389272302
     0.6111518667756916    0.4443686229305228    0.0386734975270231
     0.3888624100058851    0.5555249247852253    0.2716327121985637
     0.5554925363513302    0.2222198622500575    0.1937617168474993
     0.7778177623049597    0.1109877059174573    0.1166214455809145
     0.7776192485515546    0.1111526134584207    0.0385954894198872
     0.5555303718900877    0.2222207060256461    0.2716270173906355
     0.8888648392892371    0.5555750199251784    0.1937828129873172
     0.1110622383540729    0.4446454615576127    0.1166182604534135
     0.1113447470662672    0.4447744332475275    0.0386207479167278
     0.8888372215209998    0.5555361507452542    0.2716324074612659
     0.0555558094182941    0.2222450798616893    0.1937803841060956
     0.2775743730175488    0.1109764415022620    0.1165905114436476
     0.2778495019076344    0.1111164021861548    0.0386324870804961
     0.0555130437012722    0.2222131042000333    0.2716289669476181
     0.2222106548618766    0.8888544959623818    0.1937857384997752
     0.4444295613868955    0.7776135527830869    0.1165561103223271
     0.4443493896339342    0.7773178960985829    0.0386674532367567
     0.2221990860284071    0.8888915391806865    0.2716370754309396
     0.7221570095841615    0.8888527334117361    0.1937851520887093
     0.9446031527265070    0.7780456995945041    0.1165978909742685
     0.9445927197814954    0.7779879248737340    0.0385671901920796
     0.7221889058237265    0.8889079569657714    0.2716349064459687
     0.3881600556671903    0.5554898487458650    0.0776963160171359
     0.5553804180295341    0.2213439007893907    0.0776879845071511
 
\end{verbatim}
\end{center}

\subsubsection{25.00\%}
\label{struct:25}
\begin{center}
\begin{verbatim}
Mo{1-x}S2-W{x}S2 @ 25.00% concentration of W
   1.00000000000000     
     6.3837702599614250    0.0041338537133348    0.0000000000000000
    -1.6002842418061685    8.2895127991320852    0.0000000000000000
     0.0000000000000024    0.0000000000000030   39.9862299999999991
  Mo              S               W             
     9    24     3
Direct
     0.6110769712874340    0.4444323316461100    0.2325581561944570
     0.7777525183152251    0.1111149041687640    0.2325639308073616
     0.8897339112663720    0.5563056704857554    0.0777223518083329
     0.1111059642586225    0.4444370926369489    0.2325699829723097
     0.0546518913421084    0.2227041384732862    0.0777259688957476
     0.2777686159334207    0.1110886417536440    0.2325640314018144
     0.4444199119584383    0.7777541301583106    0.2325634917692620
     0.7234118466157592    0.8894858613406037    0.0777236100542162
     0.9444397793063019    0.7777819039503848    0.2325638929663916
     0.3888198353630944    0.5555203049229464    0.1936430530064249
     0.6113160232541147    0.4443904654520658    0.1167687314845461
     0.6110455173307981    0.4442982723671176    0.0387274159258050
     0.3888822484758407    0.5555318949116494    0.2715058174411982
     0.5554978827674808    0.2221872723791210    0.1936433849132231
     0.7778457760510111    0.1112768857952810    0.1167695909682628
     0.7776429575101405    0.1113968366482371    0.0386770737892054
     0.5555257208083501    0.2222136684557369    0.2715033569968330
     0.8888957865387763    0.5555476312160502    0.1936627456513868
     0.1108896814954495    0.4445634104601905    0.1167410193140843
     0.1111753990803974    0.4446577682039419    0.0387120396407026
     0.8888405514470528    0.5555407768245515    0.2715051997688973
     0.0555583507393251    0.2222173899868091    0.1936621893714921
     0.2775308837234037    0.1109940983336654    0.1167373691469711
     0.2777214949531822    0.1108791989953627    0.0387645243786767
     0.0555321844148134    0.2221913551461604    0.2715078892675971
     0.2222264436321224    0.8888551274406137    0.1936439196721058
     0.4446906356475002    0.7775740668066614    0.1167342335888214
     0.4444294557950880    0.7774568381540073    0.0387683936966365
     0.2222050917417605    0.8888621409310423    0.2715053506749576
     0.7221771491437732    0.8888844329599848    0.1936614688457846
     0.9445834261311188    0.7780200565784250    0.1167770577356628
     0.9448536223165223    0.7781401148005590    0.0386686598198008
     0.7221926788701928    0.8888750892049854    0.2715053135399523
     0.3879504329295231    0.5548029702075713    0.0778048122195087
     0.5562138281921936    0.2217392662390171    0.0778008036181984
     0.2213955320694161    0.8882779919972149    0.0778031686583814
 
\end{verbatim}
\end{center}

\subsubsection{33.33\%}
\label{struct:33}
\begin{center}
\begin{verbatim}
Mo{1-x}S2-W{x}S2 @ 33.33% concentration of W
   1.00000000000000     
     6.3821679248832215    0.0038266463237940    0.0000000000000000
    -1.5995326227009050    8.2890824304435089    0.0000000000000000
     0.0000000000000024    0.0000000000000030   39.9862299999999991
  Mo              S               W             
     8    24     4
Direct
     0.6110616018668776    0.4444360938197956    0.2324479239827038
     0.7777458208056959    0.1111219271457405    0.2324398675343943
     0.8889939104500407    0.5565432166816472    0.0778290754882272
     0.1110842673897483    0.4444240300607549    0.2324542335315367
     0.2777449670646160    0.1110945583880746    0.2324395627567570
     0.4443956034103425    0.7777573740542720    0.2324465216559126
     0.7230442664934714    0.8887966449453182    0.0778234710762236
     0.9444210265923784    0.7777756524560830    0.2324530606568524
     0.3888027569666572    0.5555100415088148    0.1935314367511036
     0.6113754823935481    0.4446890305758043    0.1169314326253375
     0.6110975787136371    0.4445444789952317    0.0387750475817614
     0.3888530482509696    0.5555449839184529    0.2713958755840332
     0.5554959217033253    0.2222408350509809    0.1935243908856429
     0.7778580492256365    0.1112354968014690    0.1169336263766114
     0.7779360148669644    0.1115301171146120    0.0387750136710537
     0.5555037657301606    0.2221997270274148    0.2713865139155979
     0.8888771731907695    0.5555434613123602    0.1935512933194312
     0.1108682076441043    0.4445907826211787    0.1169021096015875
     0.1110695969805717    0.4444236533840344    0.0388103310136411
     0.8888357771619443    0.5555385312508703    0.2713978793173979
     0.0555516012555435    0.2222392349529520    0.1935250609501612
     0.2778199194055680    0.1109185868596455    0.1169057809347112
     0.2778294915933126    0.1109798485719282    0.0388567292516782
     0.0555152364264609    0.2221803929997093    0.2713888899620538
     0.2221708293051164    0.8888476758698473    0.1935274777882559
     0.4446092305903183    0.7774465157932465    0.1168820853790999
     0.4443355716036045    0.7773162004807838    0.0388101793461094
     0.2222010422130109    0.8888918823492480    0.2713918947996755
     0.7221833137293018    0.8888776596556569    0.1935465897736810
     0.9444138123469807    0.7778690756475278    0.1168944420423184
     0.9446907276658355    0.7779589355327009    0.0387474566503176
     0.7221697693364869    0.8889025631840999    0.2713916122322001
     0.3887887263737748    0.5551067767501481    0.0779075671897507
     0.5565481674723545    0.2227765709263310    0.0779151034897413
     0.0549080195573523    0.2226072298024067    0.0779171422662951
     0.2211997030034244    0.8875402141515331    0.0779033207231696
 
\end{verbatim}
\end{center}

\subsubsection{41.67\%}
\label{struct:41}
\begin{center}
\begin{verbatim}
Mo{1-x}S2-W{x}S2 @ 41.67% concentration of W
   1.00000000000000     
     6.3800841311496130    0.0032492060791528    0.0000000000000000
    -1.5992168101360993    8.2873238373579827    0.0000000000000000
     0.0000000000000024    0.0000000000000030   39.9862299999999919
  Mo              S               W             
     7    24     5
Direct
     0.6110925642757895    0.4444467604268182    0.2323330324367779
     0.7777656830344104    0.1111188644195914    0.2323376146902461
     0.1110974224894344    0.4444339395809017    0.2323390964360971
     0.2777598405660626    0.1111019660406272    0.2323314016467780
     0.4444169168971955    0.7777654643827735    0.2323457144445804
     0.7222502868330297    0.8890449517595711    0.0779216787804098
     0.9444354413169324    0.7777658072217011    0.2323459201168605
     0.3888430161258704    0.5555655189112011    0.1934187214371157
     0.6112331098018932    0.4446912259358797    0.1170935033641825
     0.6112298255973474    0.4447229677553337    0.0388550938220936
     0.3888748606856678    0.5555317036451327    0.2712913582610997
     0.5555471600410091    0.2222355448809957    0.1934149941856436
     0.7777190033590213    0.1110772259707599    0.1170543445385344
     0.7778006528969236    0.1113423033713393    0.0388332216430744
     0.5555262823367602    0.2222266749013216    0.2712847120589649
     0.8889096347001656    0.5555631319339238    0.1934198090404453
     0.1110920796766342    0.4444695459637170    0.1170739080765804
     0.1111057998026598    0.4444772841725608    0.0388807503346763
     0.8888596976337202    0.5555290181971447    0.2712910140669904
     0.0555407542194951    0.2222331203813886    0.1934148351348555
     0.2777934075099364    0.1108384704107764    0.1170416691898005
     0.2777922942817855    0.1108897379925040    0.0388936790959518
     0.0555524705181795    0.2222087700827058    0.2712874455594549
     0.2221952205815872    0.8888371267686229    0.1934213962966266
     0.4446102748562731    0.7777801490306497    0.1170399610644660
     0.4443337624119624    0.7776011818950153    0.0388453189341764
     0.2222118503740081    0.8889058977183586    0.2712937717359158
     0.7222017416655543    0.8888702837625411    0.1934408381828163
     0.9443148727420980    0.7778364091840720    0.1170441201143402
     0.9445036499353380    0.7776603423591246    0.0388399007330587
     0.7222091974534460    0.8889014841655963    0.2712962387942704
     0.3890036163965867    0.5561449412974453    0.0780124832297531
     0.5563789520351061    0.2221892486206955    0.0780055487317840
     0.8891308341228042    0.5562537476221878    0.0780105149301287
     0.0546726527237044    0.2218859784216070    0.0780089604569923
     0.2219951685278972    0.8878532114398540    0.0779974284644647
 
\end{verbatim}
\end{center}

\subsubsection{75.00\%}
\label{struct:75}
\begin{center}
\begin{verbatim}
Mo{1-x}S2-W{x}S2 @ 75.00% concentration of W
   1.00000000000000     
     6.3788472600563804    0.0022496184732655    0.0000000000000000
    -1.5976029982928901    8.2875713212934237    0.0000000000000000
     0.0000000000000025    0.0000000000000030   40.1103900000000024
  Mo              S               W             
     3    24     9
Direct
     0.3887046178529219    0.5548647824182505    0.0788116488260868
     0.5554732034419464    0.2223552677358028    0.0788028760737549
     0.0557068477119900    0.2222108202832526    0.0788049096939610
     0.3888730100546314    0.5555616681473617    0.1926456127591904
     0.6109885962078196    0.4442005226072501    0.1177169882214356
     0.6113397159335364    0.4443834308253007    0.0398872849886250
     0.3888768781905551    0.5555807747290971    0.2706567580936711
     0.5555623281514888    0.2221782273814632    0.1926561917321641
     0.7777705774625345    0.1113333477676239    0.1176859685132229
     0.7776714150297088    0.1108767838307845    0.0399444569552780
     0.5555346934647039    0.2222314455571023    0.2706692427799920
     0.8888675553764821    0.5555348886034748    0.1926259839206136
     0.1111143974125769    0.4441170539747942    0.1177124131818991
     0.1108560006318548    0.4442966218791072    0.0398932668758118
     0.8888813600538491    0.5555886222990786    0.2706550772315624
     0.0554941913131427    0.2221779365457559    0.1926548331671431
     0.2779645874556234    0.1114541702890386    0.1177305595037958
     0.2778440216302019    0.1110279090098597    0.0398908163907829
     0.0555478325759446    0.2222338712213415    0.2706695249380696
     0.2222236138345117    0.8889350832202927    0.1926245681404211
     0.4443939367952408    0.7777535736766875    0.1177694435322409
     0.4445102044057119    0.7781435830313341    0.0399517838035245
     0.2221795227680285    0.8888506032531071    0.2706536197818154
     0.7222110730958864    0.8889372578777320    0.1926236538223459
     0.9445283729555594    0.7777871666219497    0.1177747830780000
     0.9445432965444965    0.7779130144615589    0.0400086011494645
     0.7222073279204175    0.8888337314904146    0.2706564949177915
     0.6111080455138513    0.4444407741988711    0.2316038646299597
     0.7777568098128507    0.1110752986109834    0.2316165811911952
     0.8885588309715795    0.5540954374381204    0.0789108681466078
     0.1110786579283695    0.4444391793506660    0.2316034224931586
     0.2777515458297894    0.1110895816849819    0.2316107766085380
     0.2233868802442315    0.8901143360058431    0.0789358924352701
     0.4444332172446354    0.7777887561620679    0.2315938006834628
     0.7216338006117985    0.8898239123762088    0.0789403541383495
     0.9444230335715005    0.7777705654334557    0.2315870777952130
 
\end{verbatim}
\end{center}

\section{High-symmetry path phonon band structures}
\label{sec:bs}

We report in \autoref{fig:bs_all} the phonon band structures of TMD homobilayers and doped heterostructures, calculated using \textsc{phonopy} along a conventional high-symmetry path for hexagonal lattices.
None of the systems exhibit imaginary frequencies, indicating that the optimised geometries are dynamically stable.

\begin{figure*}[t]
\centering

\begin{subfigure}{0.49\textwidth}
\includegraphics[width=\linewidth]{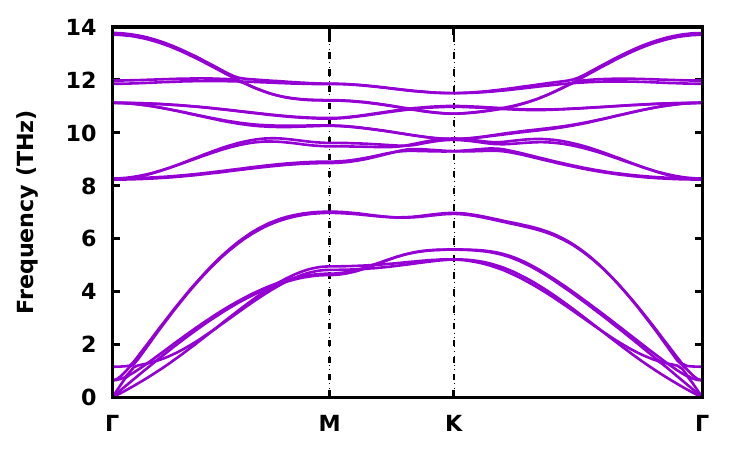}
\caption{0.00\% W}
\end{subfigure}
\hfill
\begin{subfigure}{0.49\textwidth}
\includegraphics[width=\linewidth]{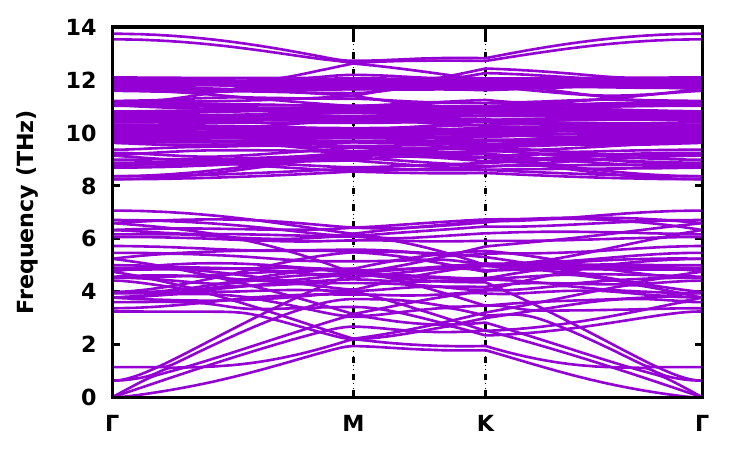}
\caption{16.67\% W}
\end{subfigure}

\vspace{0.5em}

\begin{subfigure}{0.49\textwidth}
\includegraphics[width=\linewidth]{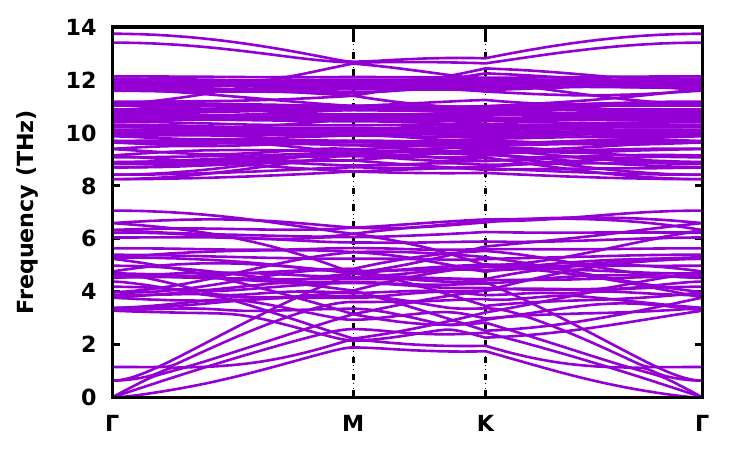}
\caption{25.00\% W}
\end{subfigure}
\hfill
\begin{subfigure}{0.49\textwidth}
\includegraphics[width=\linewidth]{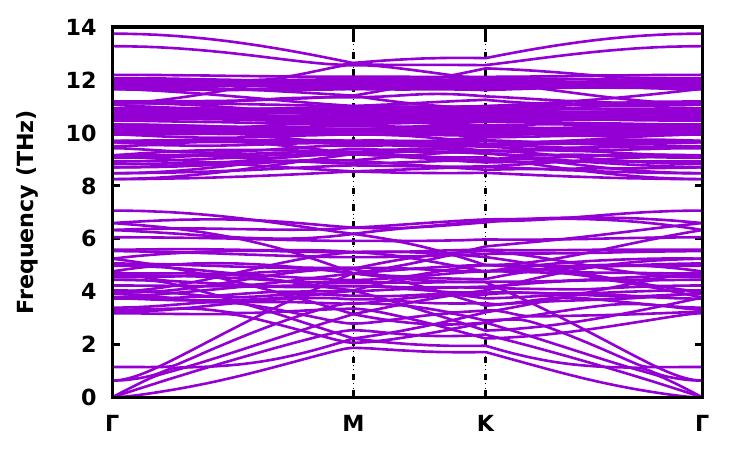}
\caption{33.33\% W}
\end{subfigure}

\vspace{0.5em}

\begin{subfigure}{0.49\textwidth}
\includegraphics[width=\linewidth]{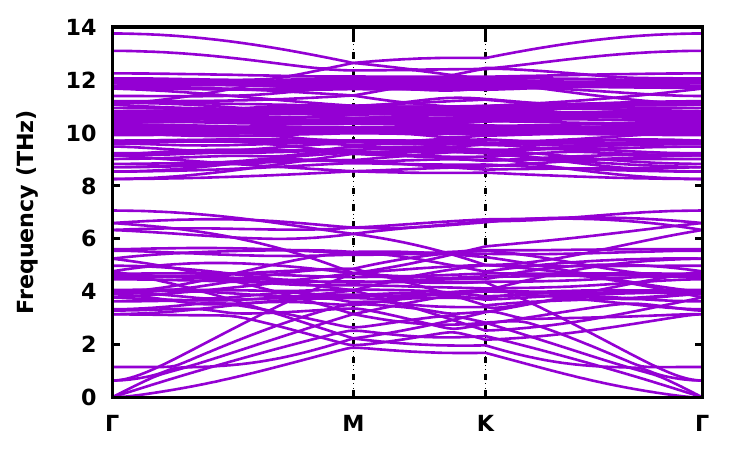}
\caption{41.67\% W}
\end{subfigure}
\hfill
\begin{subfigure}{0.49\textwidth}
\includegraphics[width=\linewidth]{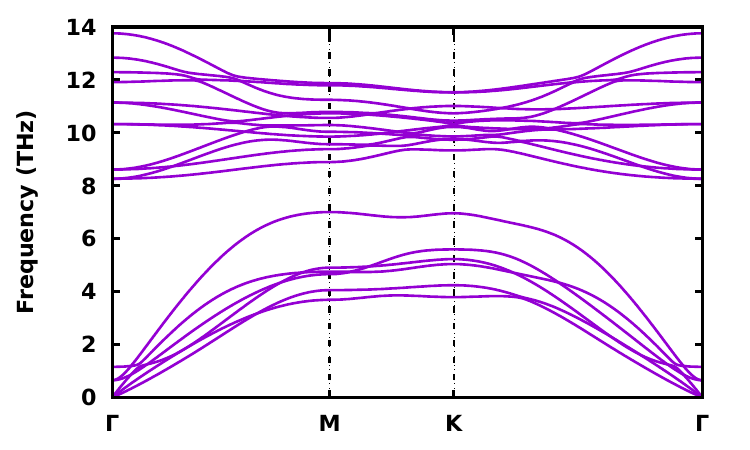}
\caption{50.00\% W}
\end{subfigure}

\vspace{0.5em}

\begin{subfigure}{0.49\textwidth}
\includegraphics[width=\linewidth]{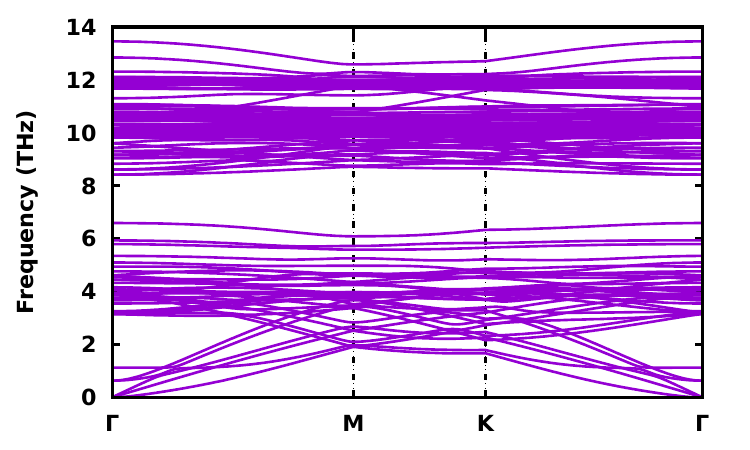}
\caption{75.00\% W}
\end{subfigure}
\hfill
\begin{subfigure}{0.49\textwidth}
\includegraphics[width=\linewidth]{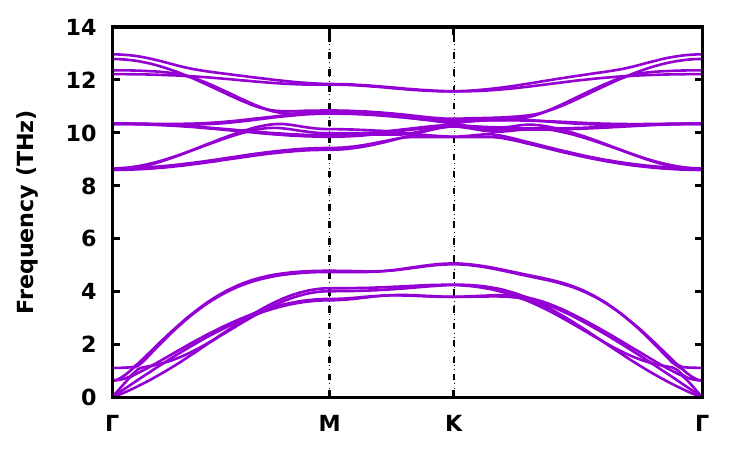}
\caption{100\% W}
\end{subfigure}

\caption{Phonon band structures along the $\Gamma$–M–K–$\Gamma$ high-symmetry path across the Mo$_{1-x}$W$_x$S$_2$ bilayer configurational landscape, where $x$ denotes the W concentration (atomic fraction). The W concentration increases from 0\% to 100\% (left to right, top to bottom).}
\label{fig:bs_all}
\end{figure*}
\end{document}